\documentclass[journal]{IEEEtran}
\usepackage{amsmath,amsfonts}
\usepackage{algorithmicx}
\usepackage{algorithm}
\usepackage{algpseudocode}
\usepackage{array}
\usepackage{textcomp}
\usepackage{stfloats}
\usepackage{url}
\usepackage{verbatim}
\usepackage{graphicx}
\usepackage{cite}
\usepackage{multirow}
\usepackage{makecell} 
\usepackage{booktabs}

\usepackage{color}
\usepackage{bm} 
\usepackage{xcolor} 
\definecolor{myblue}{RGB}{0, 0, 255} 

\ifCLASSOPTIONcompsoc
  \usepackage[caption=false,font=normalsize,labelfont=sf,textfont=sf]{subfig}
\else
  \usepackage[caption=false,font=footnotesize]{subfig}
\fi

\begin{document}

\title{An OTFS Waveform-Based Delay-Doppler Domain Channel Measurement Method for High-Mobility Scenarios}

\author{Kaifeng Bao,
        Tao Zhou,~\IEEEmembership{Senior Member,~IEEE,}
        Chaoyi Li,
        Liu Liu,~\IEEEmembership{Member,~IEEE,}
        and Bo Ai,~\IEEEmembership{Fellow,~IEEE}

\thanks{This work was supported by the Fundamental Research Funds for the Central Universities under Grant 2025YJS033, and the National Natural Science Foundation of China under Grants 62571024, 62221001, and 62341127. \textit{(Corresponding author: Tao Zhou.)}}
\thanks{K. Bao, T. Zhou, C. Li and L. Liu are with the School of Electronic and Information Engineering, Beijing Jiaotong University, Beijing 100044, China (e-mail: 23111017@bjtu.edu.cn; taozhou@bjtu.edu.cn; 21120072@bjtu.edu.cn; liuliu@bjtu.edu.cn).}
\thanks{B. Ai is with the State Key Laboratory of Advanced Rail Autonomous Operation, Beijing, 100044, China, and also with the School of Electronic and Information Engineering, Beijing Jiaotong University, Beijing 100044, China (e-mail: boai@bjtu.edu.cn).}}

\markboth{ }%
{Shell \MakeLowercase{\textit{et al.}}: Bare Demo of IEEEtran.cls for IEEE Journals}

\maketitle

\begin{abstract}
Channel measurements are the prerequisite for applying emerging transmission technologies and designing communication systems. Conventional time or frequency domain channel measurement methods cannot directly obtain Doppler information induced by high-mobility scenarios. The channel spreading function (CSF) simultaneously captures delay and Doppler information while naturally characterizing the propagation environment in the delay-Doppler (DD) domain. However, DD domain channel measurement methods remain underexplored. This paper presents an orthogonal time frequency space (OTFS) waveform-based DD domain channel measurement method for high-mobility scenarios. A native OTFS waveform, employed as the sounding signal, is designed for the first time, and its sounding capability is comprehensively analyzed. Next, we detail the methodology of DD domain channel measurement, including synchronization and CSF estimation. To enhance measurement precision, a joint fractional delay and Doppler shift estimation algorithm is proposed, and the overall performance of the proposed method is evaluated. Subsequently, a practical DD domain channel measurement system is established, followed by system calibration and verification. Finally, DD domain channel measurements are conducted in vehicle-to-infrastructure (V2I) and vehicle-to-vehicle (V2V) scenarios. Measurement results, including the CSF and other small-scale fading characteristics, confirm the effectiveness of the proposed method and offer valuable insights for advancing research on high-mobility communications.
\end{abstract}


\section{Introduction}
\IEEEPARstart{W}{hile} the fifth-generation (5G) wireless communication system is still under deployment, the potential use cases and enabling technologies of the sixth-generation (6G) system have received significant attention from academia and industry [1]-[4]. The 6G system is envisioned to support space-air-ground-sea integrated communications, which presents substantial challenges in maintaining reliable communication quality in high-mobility scenarios, such as vehicle-to-everything (V2X), high-speed railway (HSR), unmanned aerial vehicle (UAV), and low earth orbit (LEO) satellites [5]-[7]. Advanced transmission technologies, such as massive multiple-input multiple-output (MIMO), integrated sensing and communication (ISAC), reconfigurable intelligent surface (RIS), orthogonal time frequency space (OTFS), have been proposed to support reliable communication for high-mobility scenarios in future 6G systems [8]-[12]. In high-mobility scenarios, wireless channels exhibit time-frequency double selectivity, with delay and Doppler shift varying dramatically due to increased mobility and the complexity of propagation environment.

Since the performance of transmission technologies and communication systems will ultimately be limited by the propagation channels they operate over, a detailed knowledge and an accurate characterization of wireless channels in high-mobility scenarios is crucial. Channel measurements, as the most direct means of obtaining propagation characteristics, serve as a prerequisite for both the application of emerging transmission technologies and the design of communication systems in high-mobility scenarios. Conventional channel measurement methods are typically categorized as either active or passive, depending on whether a dedicated sounding signal is transmitted in the measurement scenario. Active measurement methods involve transmitting a dedicated signal to excite the channel and receiving the response for further analysis. Passive measurement methods analyze the wireless channel based on existing signals in the wireless network.

To date, both active and passive channel measurement methods have been widely employed in high-mobility scenarios. Generally, these methods can be categorized into time domain, frequency domain, and the emerging delay-Doppler (DD) domain, as outlined below.

\subsubsection{Time domain channel measurement methods} In the time domain, wireless channels are characterized by the channel impulse response (CIR), which explicitly details the multipath delay structure. Based on the convolution relationship between the channel and the transmitted signal, measurement methods in this domain typically employ sounding signals with favorable autocorrelation properties, while the receiver (RX) extracts the CIR via a sliding correlation operation. Commonly used time domain sounding signals include pseudo-noise (PN) sequences [13], Golay sequences [14], Zadoff-Chu (ZC) sequences [15], as well as wideband code division multiple access (WCDMA) signals [16]. In [13], the Propsound channel sounder utilizing PN sequences was employed to extract the CIRs in the HSR viaduct scenario, effectively parameterizing small-scale fading characteristics, including Doppler shifts. Beyond dedicated active signals, a WCDMA signal-based passive channel measurement method was applied to characterize time dispersion properties in diverse HSR scenarios, including plain, U-shape cutting, station, and hilly terrain [16]. Notably, since the CIR primarily depicts the multipath delay information, Doppler shifts are typically acquired indirectly by applying a Fourier transform to the temporal correlation function of instantaneous continuous CIRs, or through parameter estimation algorithms, such as space-alternating generalized expectation-maximization (SAGE) [17].

\subsubsection{Frequency domain channel measurement methods} In the frequency domain, wireless channels are characterized by the channel transfer function (CTF), which describes the frequency-selective fading properties. Based on the relationship between the time and frequency domains, the RX extracts the CTF directly by frequency domain correlation, while the CIR is subsequently obtained via the inverse Fourier transform. Orthogonal frequency-division multiplexing (OFDM) [18], [19] and other multi-carrier signals [20], [21] have been employed to estimate the CTF in various V2X communication scenarios, including highways, urban areas, tunnels, and large vehicle obstructions. Based on the measured CTF, key channel characteristic parameters, including non-stationary, large-scale and small-scale parameters, were extracted. Moreover, because of the measurement restriction and efficiency issues of applying conventional channel sounders in HSR scenarios, a passive measurement method was proposed to extract the CTF from the long-term evolution (LTE) signal transmitted in the railway network [22]. Based on 5G signals, a passive channel measurement method was proposed in [23], where channel parameters including the root-mean-square (RMS) delay spread (DS) and K-factor (KF) were derived to characterize HSR scenarios. Since the CTF does not natively capture Doppler information, Doppler shifts are typically inferred from the indirectly obtained CIRs.

\subsubsection{DD domain channel measurement methods}
In the DD domain, wireless channels are characterized by the channel spreading function (CSF), which provides a direct physical interpretation of the propagation environment. To date, mainstream DD domain channel measurement methods are predominantly built upon conventional time or frequency domain approaches, where the CSF is indirectly derived from the measured CIR or CTF via further mathematical transformations. For instance, the CSF in a V2X urban scenario was characterized using the complex Fourier transform of the CIR [24]. In [25], the sparsity of the CSF in vehicle-to-infrastructure (V2I) millimeter-wave urban street scenarios was described using the symplectic finite Fourier transform (SFFT) of the CTF. Furthermore, the sparsity and compactness of the CSF in HSR viaduct and tunnel scenarios were analyzed, with the SFFT applied to the CTFs under different DD domain grid sizes [26]. Notably, delay and Doppler shifts can be directly extracted from the obtained CSF.

\begin{table*}[t]
\centering
\caption{Comparison of Existing Time, Frequency and DD Domain Channel Measurement Methods}
\label{tab:comparison_methods}
\renewcommand{\arraystretch}{1.4} 
\small 
\begin{tabular}{m{1.0cm} | m{3.6cm} | m{4.0cm} | m{3.8cm} | m{3.5cm}}
\hline \hline

\textbf{Domain} & \textbf{Channel Representation} & \textbf{Channel Response Extraction} & \textbf{Common Sounding Signals} & \textbf{Doppler Shift Estimation} \\
\hline

\textbf{Time} & 
\textbf{CIR}: Explicitly details the multipath delay structure. & 
Extracted via \textbf{sliding correlation}. &  
\textbf{Active}: PN, Golay, ZC seqs. \par \textbf{Passive}: WCDMA signals. & 
\multirow{2}{3.6cm}{\textbf{Indirect}: Resolved via Fourier transform of the correlation of CIRs, or parameter estimation algorithms (e.g., SAGE).} \\
\cline{1-4}

\textbf{Freq.} & 
\textbf{CTF}: Describes frequency-selective fading properties. & 
Extracted via \textbf{frequency domain correlation}. &
\textbf{Active}: OFDM and other multi-carrier signals. \par \textbf{Passive}: LTE and 5G signals. & 
\\
\cline{1-5}

\textbf{DD} & 
\textbf{CSF}: Provides direct physical interpretation, featuring separability, sparsity, compactness, and stability [27]. & 
\textbf{Indirectly} extracted from measured CIR or CTF. & 
Relies on conventional time or frequency domain sounding signals. & 
\textbf{Direct}: Delay and Doppler shifts are {simultaneously extracted} from CSF. \\

\hline
\end{tabular}
\end{table*}

To comprehensively compare the existing time, frequency, and DD domain channel measurement methods, Table I summarizes their key differences in terms of channel representations, channel response extraction, common sounding signals, and Doppler shift estimation. Specifically, the CSF exhibits superior separability, sparsity, compactness, and stability compared to the time domain CIR and frequency domain CTF, making it a robust representation of high-mobility channels [27]. Furthermore, the CSF enables the direct and simultaneous extraction of both delay and Doppler shifts, rather than the indirect estimation in traditional methods. 
Despite the inherent advantages of DD domain channel measurement methods for high-mobility scenarios, current DD domain approaches still rely on conventional time or frequency domain channel sounding signals, meaning that the CSF is indirectly obtained from the measured CIR or CTF. 
For time domain channel measurement methods, stringent sampling rate requirements arise from the high chip rate of wideband sounding signals, thereby increasing system design complexity and hardware burden [28]. For frequency domain channel measurement methods, inter-carrier interference (ICI) induced by rapid channel variation under high mobility adversely affects CTF estimation [29], [30]. Consequently, existing DD domain channel measurement methods remain subject to the practical challenges associated with conventional time or frequency domain approaches. These challenges motivate the investigation of DD domain native sounding waveforms for channel measurement in high-mobility scenarios.

However, DD domain native waveforms, such as OTFS, are rarely utilized as sounding signals in existing channel measurement methods. Although OTFS pilot-aided channel estimation, OTFS sensing, and OTFS-ISAC have been extensively investigated [31]–[34], their objectives and design requirements differ from those of channel measurement. OTFS pilot-aided channel estimation mainly aims to estimate the effective channel response at the RX for subsequent signal detection and equalization, with limited emphasis on characterizing the physical propagation environment. OTFS sensing and OTFS-ISAC primarily focus on extracting target-related parameters, such as range and velocity, rather than characterizing the overall multipath propagation environment. Therefore, these OTFS-based techniques are not specifically designed for DD domain channel measurement, and a comprehensive OTFS waveform-based DD domain channel measurement framework encompassing waveform design, methodology, and practical implementation is still lacking. To address the aforementioned research gaps, this paper proposes an OTFS waveform-based DD domain channel measurement method for high-mobility scenarios. The main contributions and novelties of this work are summarized as follows.

1) A native OTFS waveform, employed as the sounding signal and strategically embedded with PN sequences, is designed for DD domain channel measurement for the first time. Furthermore, the sounding capability of this designed signal is analyzed, providing waveform design guidelines for future DD domain channel measurements.

2) The methodology for OTFS waveform-based DD domain channel measurement is detailed, encompassing synchronization and CSF estimation. To further enhance measurement precision, an algorithm for jointly estimating fractional delay and Doppler shift is proposed. Finally, the overall performance of the proposed method is comprehensively evaluated.

3) Based on the proposed method, we develop a practical DD domain channel measurement system, followed by system calibration and verification. Verification results in time-varying multipath channels confirm the reliability of the proposed method and the established system for high-mobility scenarios.

4) We conduct DD domain channel measurements in V2X urban scenarios, covering V2I and vehicle-to-vehicle (V2V) scenarios. Measurement results, including the CSF and other small-scale fading characteristics, such as power delay profile (PDP), Doppler power spectral density (DPSD), number of multipath components (MPCs), KF, RMS DS and RMS Doppler spread (DPS), are derived, which further validate the effectiveness of the proposed method.


The remainder of this paper is outlined as follows. Section II describes the waveform design for DD domain channel measurement. Section III highlights the DD domain channel measurement methodology. In Section IV, the channel measurement system is established and the channel measurement method is verified. In Section V, the measurement campaigns are described and the measurement results are analyzed. Finally, the conclusions are drawn in Section VI.

\section{Waveform Design for DD Domain Channel Measurement}
In this section, the waveform design of the sounding signal will be presented, and the sounding capability of the designed signal will be analyzed.

\begin{figure}[!t]
\center
\includegraphics[scale=0.24]{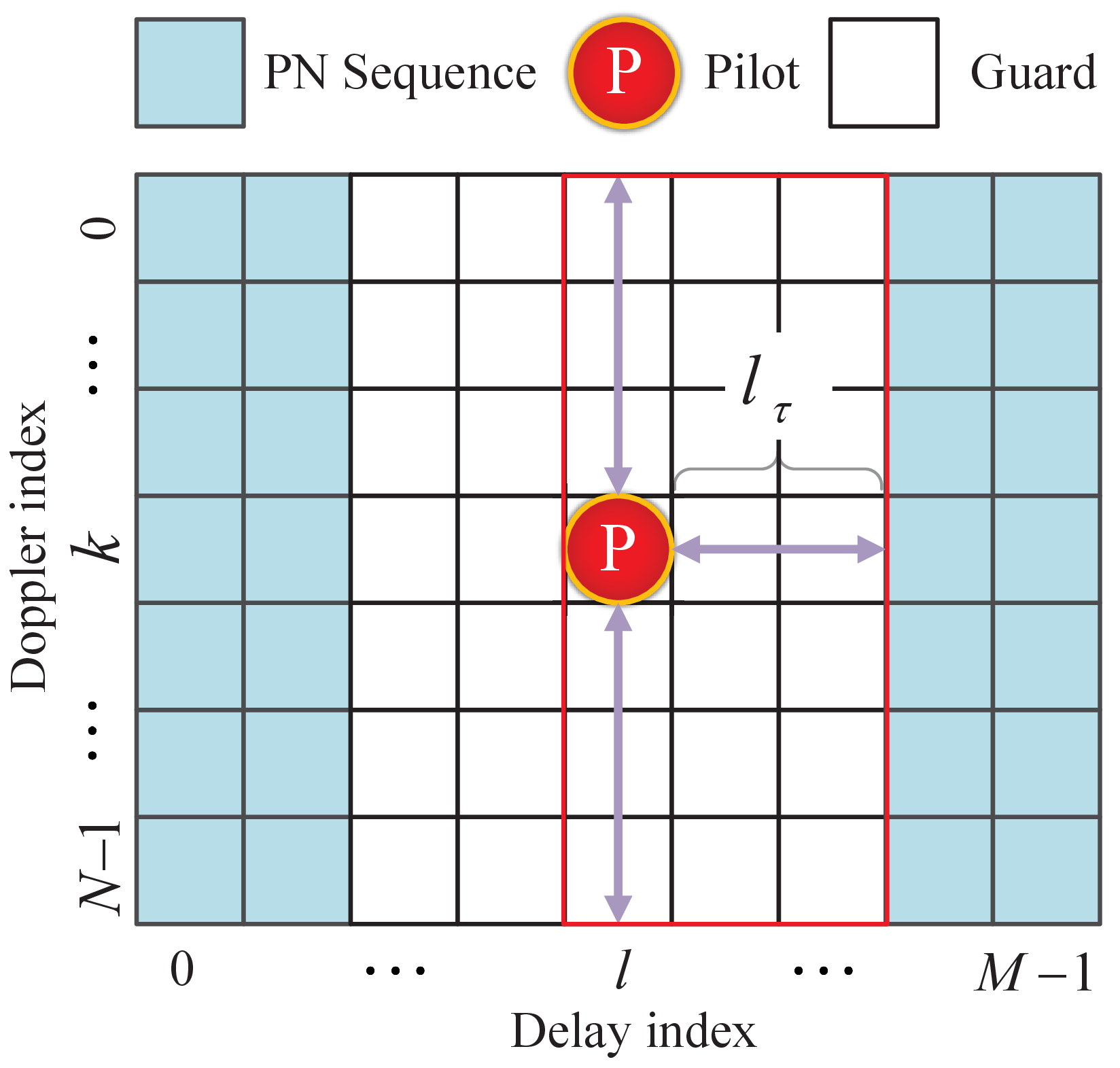}\\
\caption{Sounding signal pattern in the DD domain. \label{fig:1}}
\vspace{-0.45cm}
\end{figure}

\vspace{-0.3cm}
\subsection{Sounding Signal}
According to the basic concepts of OTFS modulation [35], [36], the waveform design of the sounding signal is performed in the DD domain. First, the time-frequency signal plane is discretized to a $N \times M$  grid ( for some integers $N$, $M$ $\textgreater$ 0 ) by sampling time and frequency axes at intervals $T$ (seconds) and $\Delta f$, respectively, i.e., $ {\Lambda}_{TF}=\left\{ \left( nT, m\Delta f \right),n=0,\cdots ,N-1, m=0,\cdots ,M-1 \right\}.$ The modulated time-frequency samples $X[n,m]$ are transmitted over an OTFS frame with duration ${{T}_{\text{OTFS}}}=NT$ and bandwidth $B=M\Delta f$.

Then, the DD signal plane is discretized to a $N \times M$ grid $ {\Lambda }_{DD}\!=\!{ \left\{ \left( k\Delta \upsilon ,l\Delta \tau \right),k=0,\cdots, N-1, l=0,\cdots, M-1 \right\} },$ where ${\Delta \upsilon =1}/{NT}$ and ${\Delta \tau =1}/{M\Delta f}$ represent the Doppler resolution and delay resolution, respectively.

We arrange the pilot, guard symbols, and PN sequences in the DD domain grid to implement OTFS waveform as the sounding signal, as depicted in Fig. 1. The pilot is placed at the center of the DD domain grid and is surrounded by guard symbols. Aside from the pilot and guard symbols, data symbols for communication are typically inserted at the remaining positions in the DD domain grid, as shown in [36]. From the perspective of channel measurement, we inserted PN sequences at the remaining positions instead of data symbols. Therefore, the sounding signal in the DD domain can be written as
\begin{equation}\label{x_kl}
  x\left[ k,l \right]=\left\{ \begin{array}{*{35}{l}}
   1 & k={{k}_{p}},l={{l}_{p}},  \\
   0 & 0\le k\le N-1,{{l}_{p}}-{{l}_{\tau }}\le l\le {{l}_{p}}+{{l}_{\tau }},  \\
   \pm {{A}_{\text{PN}}} & \text{otherwise,}  \\
\end{array} \right.
\end{equation}
where $x[k,l]$ is the symbol of the $k$-th Doppler tap and the $l$-th delay tap of the grid ${\Lambda }_{DD}$, and $\left( k_p, l_p \right )$  is the grid location corresponding to the pilot. When the pilot is laid at the center of the grid, $k_p = N/2$, $l_p = M/2$. The red dashed box in Fig. 1 indicates the DD domain grid range used to extract the CSF, i.e.,
$0\le k\le N-1$, ${{l}_{p}}\le l\le {{l}_{p}}+{{l}_{\tau }}$, where ${l}_{\tau }$ 
denotes the taps corresponding to the maximum measurable delay of the sounding signal. And ${A}_{\text{PN}}$ represents the amplitude of the PN sequences.

Since the favorable autocorrelation properties of PN sequences, the designed sounding signal exhibits excellent autocorrelation characteristics when transformed from the DD domain to the time domain. Therefore, the designed signal can be used as a synchronization signal, enabling the implementation of a high-accuracy synchronization method for DD domain channel measurement. Furthermore, the insertion of the PN sequences effectively reduces the peak-to-average power ratio (PAPR) of the sounding signal, thereby alleviating the hardware implementation burden. Detailed analyses of the synchronization method and PAPR performance are provided in Section III.

It should be noted that the inserted PN sequences may introduce interference into the CSF extraction region under non-ideal conditions, including fractional delay and Doppler shifts, non-ideal pulse shaping, and synchronization errors. When the transmit and receive pulses approximately satisfy the biorthogonality condition and the channel satisfies the underspread assumption, the influence of pulse shaping on DD domain leakage is relatively small, with the leakage mainly associated with fractional delay and Doppler shifts [37]. Fractional Doppler shift mainly causes PN sequence leakage along the Doppler dimension. However, since the CSF extraction region contains no PN sequences along this dimension, such leakage does not introduce additional interference into this region. In contrast, fractional delay may cause PN energy to leak across the delay boundary and affect the edge of the CSF extraction region. Nevertheless, even for a severe fractional offset of 0.5, the leakage power rapidly decreases and is approximately 30 dB lower than the peak at a distance of ten DD domain grid points [36]. In addition, synchronization errors may shift the received symbols in the DD domain, while fractional timing errors may introduce delay domain leakage similar to that caused by fractional delay. Therefore, with an appropriate setting of $l_{\tau}$ and accurate synchronization, interference from the PN sequences is mainly confined to the edge of the CSF extraction region and has limited influence on CSF extraction and subsequent MPC parameter estimation.

To convert the sounding signal to the time domain for transmission, the inverse symplectic finite Fourier transform (ISFFT) is first applied to the DD domain sounding signal $x[k,l]$ to obtain the time-frequency domain sounding signal $X[n,m]$. Next, the Heisenberg transform is applied to $X[n,m]$ using transmit pulse $g_{tx}(t)$ to generate the time domain sounding signal $x(t)$.

\vspace{-0.3cm}
\subsection{Sounding Capability Analysis}
The sounding capability of the designed signal can be evaluated according to several key metrics, including delay resolution, Doppler resolution, maximum measurable delay, maximum measurable Doppler shift, and minimum measurable stationary interval (SI). Table II presents a comparative analysis of these metrics under representative configurations of OTFS frame sizes $(M,N)$, and bandwidths $B$ when $l_\tau = M/4$ is set. Based on the parameters detailed in Table II, the sounding capability is analyzed as follows.

\begin{table*}[t]  
\begin{center}
\renewcommand\arraystretch{1.2}
\renewcommand{\a}{5.5}  
\small \caption{Main parameters of the sounding signal} \label{table1} \centering
\vspace{-0.2cm}
\begin{tabular}{m{\a cm}<{\centering}|c|c|c|c|c|c}
\hline\hline
\textbf{Main Parameters}&\multicolumn{6}{c}{\textbf{Values}} \\
\hline
Size of an OTFS frame $(M$, $N)$ & (4096, 2048)  & (2048, 256) & (1024, 512) & (1024, 512) & (512, 256) & (256, 128)\\
\hline
Bandwidth [MHz]  & 100 & 100 & 100 & 50 & 50 & 50\\
\hline
Frame length [ms] & 83.89 & 5.24  & 5.24 & 10.49 & 2.62  & 0.66 \\
\hline
Delay resolution [ns] & 10  & 10  & 10  & 20  & 20  & 20 \\
\hline
Doppler resolution [Hz]  & 11.92  & 190.73  & 190.73  & 95.37  & 381.47  & 1525.88 \\
\hline
Maximum measurable delay [$\mu$s] & 10.24  & 5.12  & 2.56 & 5.12 & 2.56  & 1.28 \\
\hline
Maximum measurable Doppler shift [kHz] & 12.21  & 24.41 & 48.83 & 24.41  & 48.83  & 97.66 \\
\hline
Minimum measurable SI [ms] & 83.89 & 5.24  & 5.24 & 10.49 & 2.62  & 0.66 \\
\hline
\end{tabular}
\vspace{-0.45cm}
\end{center}
\end{table*}

\subsubsection{Delay Resolution}
The delay resolution ${\Delta \tau}$  represents the ability to distinguish multipath delays, calculated as
\begin{equation}\label{delta_tau}
  {\Delta \tau =1}/{B},
\end{equation}
where $B$ denotes the bandwidth of the sounding signal. A wider bandwidth leads to improved delay resolution. For instance, the 100 MHz and 50 MHz bandwidths listed in Table II yield delay resolutions of 10 ns and 20 ns, respectively. Adequate delay resolution is fundamental to ensuring that MPCs originating from closely spaced reflectors are separable along the delay dimension of the DD domain grid.

\subsubsection{Doppler Resolution}
The Doppler resolution ${\Delta \upsilon}$ quantifies the ability to distinguish Doppler shifts, calculated as
\begin{equation}\label{delta_nu}
  {\Delta \upsilon = 1}/{T}_{\text{OTFS}},
\end{equation}
where ${T}_{\text{OTFS}}$ represents the frame length of the sounding signal. As indicated in Table II, a configuration with a short frame length (e.g., $\left( 256, 128 \right)$ at 50 MHz) results in a coarse Doppler resolution, which is inadequate for effective channel measurements. The Doppler resolution can be refined by either employing a larger frame size or a narrower bandwidth. Specifically, under a fixed bandwidth, expanding the frame size leads to finer resolution, as evidenced by the 11.92 Hz achieved with the $\left( 4096, 2048 \right)$ and 100 MHz configuration. Alternatively, for a fixed frame size, such as $\left( 1024, 512 \right)$, reducing the bandwidth from 100 MHz to 50 MHz refines ${\Delta \upsilon}$ from 190.73 Hz to 95.37 Hz.

\subsubsection{Maximum Measurable Delay}
The maximum measurable delay ${\tau_{max}}$ of the sounding signal can be calculated as

\begin{equation}\label{tau_max}
  {\tau_{max}} = l_{\tau} \Delta \tau.
\end{equation}
Therefore, $l_{\tau}$ directly determines the maximum measurable delay for a given bandwidth. A larger ${\tau}_{max}$ can be achieved by increasing $l_{\tau}$, which correspondingly enlarges the guard region and reduces the number of PN sequences. The setting $l_{\tau} = M/4$ is adopted in Table II, yielding ${\tau_{max}} =  M/ \left(4B\right).$
The $\tau_{max}$ must be sufficiently large to capture the MPCs that may exist in the wireless channel. In high-mobility scenarios, the propagation delay difference of MPCs can reach up to 2 $\mu s$. Consequently, for a 50 MHz bandwidth, the number of delay taps $M$ should exceed 400. As shown in Table II, the configuration with $\left( 256, 128 \right)$ and 50 MHz provides a $\tau_{max}$ of only 1.28 $\mu s$, which is insufficient to capture MPCs with large delays. Moreover, $\tau_{max}$ can be extended by either reducing the bandwidth $B$ or by increasing the delay taps $M$. For instance, comparing the $\left( 2048, 256 \right)$ and $\left( 1024, 512 \right)$ configurations under a fixed 100 MHz bandwidth, the former provides a larger $\tau_{max}$ while maintaining identical resolution performance.

\subsubsection{Maximum Measurable Doppler Shift}
The maximum measurable Doppler shift ${\upsilon_{max}}$ of the sounding signal can be calculated as
\begin{equation}\label{nu_max}
  {{\upsilon }_{\max }}=\Delta \upsilon \cdot {N}/{2}\;={B}/{\left( 2M \right)}.
\end{equation}
Accordingly, the measurable Doppler shift range in the DD domain grid is $\left[-{\upsilon }_{\max }, {\upsilon }_{\max }\right)$.
It can be seen that the ${\upsilon_{max}}$ is jointly determined by the bandwidth $B$ and the number of delay taps $M$, where $B$ is typically much larger than $M$. Therefore, as shown in Table II, ${\upsilon_{max}}$ is usually on the order of kilohertz, which is sufficient to meet the measurement requirements in most high-mobility scenarios.

\subsubsection{Minimum Measurable SI}
The minimum measurable SI $SI_{min}$ of the sounding signal can be determined as
\begin{equation}\label{SI_min}
  SI_{min} = {T}_{\text{OTFS}}.
\end{equation}
To ensure the validity of channel measurements, $SI_{min}$ must not exceed the actual SI of the wireless channel. The SI varies across different scenarios due to the dynamic evolution of MPCs in the time domain [38]. Typically, the SI is on the order of seconds [39], but it decreases to the millisecond level in high-mobility scenarios such as V2X [39] and HSR [40]. Consequently, for large OTFS frame sizes like $\left( 4096, 2048 \right)$ which correspond to a frame length of 83.89 ms, the resulting $SI_{min}$ becomes excessively long, thereby rendering effective DD domain channel measurement unfeasible in high-mobility scenarios.

In summary, the waveform design for DD domain channel measurements necessitates a comprehensive consideration of the aforementioned metrics. To ensure the validity of channel measurements, the $SI_{min}$ must be strictly smaller than the actual SI of the target scenario. Under this condition, a larger OTFS frame size should be selected to enhance the channel sounding capability. Furthermore, since the continuously varying delays and Doppler shifts in realistic channels rarely align perfectly with the discrete DD domain grids, fractional delay and Doppler effects are inevitable, even with sufficient delay and Doppler resolution. To address this issue, a joint fractional delay and Doppler shift estimation algorithm is proposed to further improve measurement precision. The specific details of this algorithm will be elaborated in Section III.
\section{DD Domain Channel Measurement Methodology}
In this section, the principles of synchronization and CSF estimation within the proposed DD domain channel measurement framework are described. Subsequently, a joint fractional delay and Doppler shift estimation algorithm is presented to enhance measurement precision. Finally, the performance of the proposed method is comprehensively analyzed.

\subsection{Synchronization}
The sounding signal $x(t)$ is transmitted over the wireless channel and subsequently captured by the RX. At the RX, synchronization is a prerequisite for accurately recovering the DD domain signal from the received time domain signal $y(t)$. Existing synchronization methods for OTFS systems can be broadly classified into two categories: preamble-based and pilot characteristics-based approaches. Authors in [41] designed a random access preamble and developed a timing offset (TO) estimation method for the OTFS uplink system. In [42], a linear frequency modulated signal is employed as a preamble for downlink synchronization. In DD domain channel measurements, although inserting a preamble before the sounding signal can enhance synchronization performance, it inevitably extends the frame length, thereby increasing the $SI_{min}$. In contrast, pilot characteristics-based methods exploit the periodic properties of pilots in the time-delay domain, enabling synchronization without additional overhead or increasing $SI_{min}$ [43], [44]. Based on this principle, a maximum length sequence is embedded in the time-delay domain to enhance synchronization performance in OTFS systems [45]. While pilot characteristics-based synchronization avoids additional overhead, it relies exclusively on the pilot, leaving the positions in the remainder of the OTFS frame unutilized.

To fully exploit the OTFS frame, the positions of the data symbols typically used for communication are occupied by the PN sequences in the DD domain for the purpose of channel measurement. This design endows the sounding signal with superior correlation properties in the time domain, which facilitates synchronization using a sliding correlation approach without increasing the $SI_{min}$. Furthermore, to improve the computational efficiency of the sliding correlation, only the initial portion of the sounding signal frame can be selected as the synchronization signal, rather than the entire frame. This optimization markedly reduces processing complexity while preserving sufficient synchronization performance.

During DD domain channel measurement, the transmitter (TX) continuously transmits the sounding signal, while the RX finds the peak position through the sliding correlation function between the received signal and the local synchronization signal, thereby determining the starting position of each OTFS frame to achieve synchronization. The sliding correlation function is defined as follows
\begin{equation}\label{corr_fun}
  {{R}_{c}}\left[ k \right]={{\left| \sum\limits_{i=0}^{L-1}{y\left[ k+i \right]\cdot {{s}^{*}}\left[ i \right]} \right|}^{2}},
\end{equation}
where $k = 0,1,\cdots, MN-1 $ and ${\left( \cdot \right)} ^ {*}$ denotes the conjugation operation. Here, $y[i]$ and $s[i]$ denote the discrete representations of the received signal $y(t)$ and the synchronization signal, respectively. Let $L$ denote the length of the synchronization signal. Specifically, the case $L=MN$ corresponds to utilizing the entire frame of the sounding signal, whereas $L \textless MN$ implies that only the first $L$ samples are used. Consequently, the starting position $k_{\text{TO}}$ of an OTFS frame is determined by
\begin{equation}\label{syn_position}
  {{k}_{\text{TO}}}=\underset{0\le k\le MN-1}{\mathop{\arg \max }}\,\left\{ {{R}_{c}}\left[ k \right] \right\}.
\end{equation}
Starting from the estimated position ${k}_{\text{TO}}$, the continuously received data are successively segmented into OTFS frames, each containing $MN$ samples.

\subsection{CSF Estimation}
In the DD domain, the sparse representation of the channel is
\begin{equation}\label{h_DD_representation}
  h(\tau ,\upsilon )=\sum\limits_{i=1}^{P}{{{h}_{i}}\delta (\tau -{{\tau }_{i}})\delta }(\upsilon -{{\upsilon }_{i}}),
\end{equation}
where $P$ is the number of propagation paths, ${h}_{i}$, ${\tau}_{i}$, and ${\upsilon}_{i}$ represent the path gain, delay, and Doppler shift associated with the  $i$-th path, and $\delta( \cdot )$ denotes the Dirac delta function.

The sounding signal $x(t)$ is transmitted over the wireless channel with CSF $h(\tau ,\upsilon )$. The received signal $y(t)$, after synchronization, is first filtered with the receive pulse $g_{rx}(t)$ followed by a sampler, yielding $Y[n,m]$ in the time-frequency domain. Then, the SFFT is applied to $Y[n,m]$ for obtaining received symbols $y[k,l]$ in the DD domain [35].

The relation between $y[k,l]$ and $x[k,l]$ was derived in [36] as
\begin{equation}\label{DD_IO_relation}
  \begin{split}
      y[k,l]= &\sum\limits_{{k}'=0}^{N-1}{\sum\limits_{{l}'=0}^{M-1}{{b}\left[ {k}',{l}' \right]\cdot \hat{h}\left[ {k}',{l}' \right]x\left[ {{\left( k-{k}' \right)}_{N}},{{\left( l-{l}' \right)}_{M}} \right]}} \\
      &+n\left[ k,l \right],
  \end{split}
\end{equation}
where $n[k,l]\sim \mathcal{CN}(0,{{\sigma}^{2}})$ is additive white noise with variance ${\sigma}^{2}$, and ${\left( \cdot \right)}_{N}$, ${\left( \cdot \right)}_{M}$ denote modulo $N$ and $M$ operations, respectively. $\hat{h}\left[ {k}',{l}' \right]=h\left[ {k}',{l}' \right]{{e}^{-j2\pi \tfrac{{{k}'}}{NT}\tfrac{{{l}'}}{M\Delta f}}}$, $b\left[ {k}',{l}' \right]\in \left\{ 0,1 \right\}$ is the path indicator, i.e., $b\left[ {k}',{l}' \right] = 1$ indicates that there is a path with Doppler tap ${k}'$ and delay tap ${l}'$ with corresponding path magnitude $\hat{h}\left[ {k}',{l}' \right]$, otherwise, there is no such path, i.e., $b\left[ {k}',{l}' \right] = 0$ and $\hat{h}\left[ {k}',{l}' \right] = 0$. We have the total number of paths
\begin{equation}\label{Path_num}
  P=\sum\limits_{{k}'=0}^{N-1}{\sum\limits_{{l}'=0}^{{{l}_{\tau }}}{{b}\left[ {k}',{l}' \right]}}.
\end{equation}
Each path circularly shifts the transmitted symbols by the delay and Doppler taps [36].

Substituting the expression of $x[k,l]$ into Eq. (10), $y[k,l]$ can be rewritten as
\begin{equation}\label{y_kl}
  y[k,l]=b\left[ k-{{k}_{p}},l-{{l}_{p}} \right]\cdot \hat{h}\left[ k-{{k}_{p}},l-{{l}_{p}} \right]+n\left[ k,l \right]
\end{equation}
for $k \in \left[ 0, N-1 \right]$, $l \in \left[ l_{p}, l_{p} + l_{\tau} \right]$. We can see that if there is a path with Doppler tap $k-k_p$ and delay tap $l-l_p $, i.e., $b\left[ k-{{k}_{p}},l-{{l}_{p}} \right] = 1$, we have $ y[k,l]=\hat{h}\left[ k-{{k}_{p}},l-{{l}_{p}} \right]+n\left[ k,l \right]$. Otherwise, $ y[k,l] = n\left[ k,l \right]$  . Therefore, $y[k,l]$ for $k \in \left[ 0, N-1 \right]$, $l \in \left[ l_{p}, l_{p} + l_{\tau} \right]$ is the measured CSF. 
However, the measured CSF is inevitably affected by fractional delay and Doppler shift effects due to mismatch between the continuous channel parameters and the discrete DD domain grids. Therefore, further refinement of the CSF is required to achieve more accurate channel measurement.

\subsection{Fractional Delay and Doppler Shift Estimation}
To address the fractional delay and Doppler effects, we propose a joint fractional delay and Doppler shift estimation algorithm to improve the precision of channel measurement. The proposed algorithm operates by first constructing an equivalent channel function, followed by calculating the two-dimensional correlation function between the measured CSF and the equivalent channel function to extract MPC parameters. In the following, we derive the equivalent channel function and elaborate on the details of the proposed algorithm.

\subsubsection{Equivalent channel function}
Based on the input-output relationship of OTFS modulation [35], the received signal can be expressed in the DD domain as follows
\begin{equation}\label{DD_IO_relationship}
  \begin{split}
  y\left[ k,l \right]=
  &\frac{1}{MN}\sum\limits_{{k}'=0}^{N-1}{\sum\limits_{{l}'=0}^{M-1}{x\left[ {k}',{l}' \right]\cdot {{h}_{w}}\left( \frac{\Delta k}{NT},\frac{\Delta l}{M\Delta f} \right)}}\\
  &+n\left[ k,l \right],
  \end{split}
\end{equation}
where $n[k,l]\sim \mathcal{CN}(0,{{\sigma}^{2}})$ is additive white noise with variance ${\sigma}^{2}$, and $\Delta k = k - k'$ and $\Delta l = l - l'$ represent the Doppler and delay tap offsets in the DD domain grid, respectively. And $h_{w}\left( \cdot , \cdot \right)$ is the continuous equivalent channel function in the DD domain, which can be expressed as
\begin{equation}\label{EQCF}
  {{h}_{w}}\left( {\upsilon }',{\tau }' \right)=\int_{\upsilon }{\int_{\tau }{h\left( \tau ,\upsilon  \right)\cdot w\left( {\upsilon }'-\upsilon ,{\tau }'-\tau  \right){{e}^{-j2\pi \upsilon \tau }}}d\tau }d\upsilon,
\end{equation}
where $ w\left( {\upsilon },{\tau } \right)$ represents the filter function of the OTFS system. 
Assuming ideal window function, $ w\left( {\upsilon },{\tau } \right)$ can be simplified to
\begin{equation}\label{W_DD_con_sim}
  w\left( {\upsilon },{\tau } \right)=\sum\limits_{m=0}^{M-1}{\sum\limits_{n=0}^{N-1}{{{e}^{-j2\pi \left( \upsilon nT-\tau m\Delta f \right)}}}}.
\end{equation}
By substituting the $h \left( \tau ,\upsilon  \right)$ into the equation above, 
the continuous equivalent channel function of the $i$-th path can be expressed as
\begin{equation}\label{h_DD_sim_ith}
\begin{split}
{h_{w,i}}\left( {\upsilon ',\tau '} \right) 
    &= {h_i} w\left( {\upsilon ' - {\upsilon _i},\tau ' - {\tau _i}} \right){e^{ - j2\pi {\upsilon _i}{\tau _i}}}\\
    &= {h_i} {e^{ - j2\pi {\upsilon _i}{\tau _i}}} \times \left( {\sum\limits_{n = 0}^{N - 1} {{e^{ - j2\pi \left( {\upsilon ' - {\upsilon _i}} \right) \cdot nT}}} } \right)\\
    & \times \left( {\sum\limits_{m = 0}^{M - 1} {{e^{j2\pi \left( {\tau ' - {\tau _i}} \right) \cdot m\Delta f}}} } \right)  .
\end{split}
\end{equation}

In the OTFS system, the above equation needs to be sampled in the DD domain to obtain the discrete equivalent channel function
\begin{equation}\label{h_DD_sim_ith_dis}
\begin{split}
&{h_{eq,i}}\left( {\frac{{\Delta k}}{{NT}},\frac{{\Delta l}}{{M\Delta f}}} \right) = {\left. {{h_{w,i}}\left( {\upsilon ',\tau '} \right)} \right|_{\upsilon ' = \frac{{\Delta k}}{{NT}},\tau ' = \frac{{\Delta l}}{{M\Delta f}}}}\\
 &= {h_i} \cdot {e^{j{\varphi _i}}} \cdot \underbrace {\left( {\sum\limits_{n = 0}^{N - 1} {{e^{ - j2\pi \left( {\Delta k - {k_i}} \right)\frac{n}{N}}}} } \right)}_{{h_{eq,{\upsilon _i}}}\left[ {\Delta k} \right]} \cdot \underbrace {\left( {\sum\limits_{m = 0}^{M - 1} {{e^{j2\pi \left( {\Delta l - {l_i}} \right)\frac{m}{M}}}} } \right)}_{{h_{eq,{\tau _i}}}\left[ {\Delta l} \right]},
\end{split}
\end{equation}
where $\varphi_i = -2\pi\upsilon_i \tau_i$ represents the phase of the $i$-th path. The last two terms denote the Doppler and delay domain equivalent channel functions, respectively. The terms $l_i = \tau_i/{\Delta \tau}$ and $k_i = \upsilon_i/{\Delta \upsilon}$ represent the normalized delay and Doppler shift of the $i$-th path, which can be decomposed into integer and fractional parts, respectively
\begin{equation}\label{l_i}
  {l_i} = {{{\tau _i}} \mathord{\left/
 {\vphantom {{{\tau _i}} {\Delta \tau }}} \right.
 \kern-\nulldelimiterspace} {\Delta \tau }} = {l_{i,I}} + {l_{i,F}},
\end{equation}
\begin{equation}\label{k_i}
{k_i} = {{{\upsilon _i}} \mathord{\left/
 {\vphantom {{{\upsilon _i}} {\Delta \upsilon }}} \right.
 \kern-\nulldelimiterspace} {\Delta \upsilon }} = {k_{i,I}} + {k_{i,F}},
\end{equation}
where ${l_{i,I}}$ and ${l_{i,F}}$ represent integer delay and fractional delay, and ${k_{i,I}}$ and ${k_{i,F}}$  represent integer Doppler shift and fractional Doppler shift, respectively.

\begin{algorithm}[!t]
\caption{Joint Fractional Delay and Doppler Shift Estimation Algorithm}
    \begin{algorithmic}[1] 
        \Statex \textbf{Input:} Measured CSF $\hat{h}[k,l]$, $M$, $N$, $B$, delay estimation step $\Delta \tau_s$, Doppler estimation step $\Delta \upsilon_s$, power threshold $P_{th}$, number of MPCs ${P_{max}}$
        \Statex \textbf{Output:} MPC parameters $\hat{h_{i}},\hat{\tau_{i}},\hat{\upsilon_{i}}$ for $i \in \{1, \dots, {P_{max}} \}$            
        \For {$i = 1$ to ${P_{max}}$}
            \State $\big \langle k_{i,I},l_{i,I} \big \rangle = \mathop{\arg\max}\limits_{0 \leq k \leq N-1 , 0 \leq l \leq M-1} \left( \big|\hat{h}[k,l]\big| \right) $
            \For{$k_{i,F}= -0.5 : \Delta \upsilon_s  : 0.5$}
                \For{$l_{i,F}= -0.5 : \Delta \tau_s  : 0.5$}
                    \State  Obtain 
                        $R_{h_{EQ,i},\hat{h}}\big[k_{i,F},l_{i,F}\big] $  using Eq. (20)               
                \EndFor
            \EndFor
        \State Extract $\big \langle \hat{k}_{i,F},\hat{l}_{i,F} \big \rangle, \hat{\upsilon_{i}}, \hat{\tau_{i}}, \hat{h_{i}}, \hat{\varphi_i}$ using Eqs. (21)-(25)
        \If{$\big|\hat{h}_i\big|^2 > P_{th}$}
            \State $\hat{h}[k,l] = \hat{h}[k,l] -  \hat{h_{i}} \cdot e^{j \hat{\varphi_i}} \cdot {h}_{EQ,i}[k,l]$
        \Else
            \State \textbf{Exit}
        \EndIf
        \EndFor
    \end{algorithmic}
\end{algorithm}

\subsubsection{Joint Fractional Delay and Doppler Shift Estimation Algorithm}
In this subsection, the proposed joint fractional delay and Doppler shift estimation algorithm for accurately extracting MPC parameters, including path gain, delay, and Doppler shift, is detailed as outlined in Algorithm 1.

The algorithm initialization involves setting parameters $M$, $N$, $B$, delay estimation step $\Delta \tau_s$, Doppler estimation step $\Delta \upsilon_s$, the power threshold $P_{th}$ and the maximum number of MPCs $P_{max}$. The threshold $P_{th}$ is predetermined to distinguish valid MPCs from the noise. In step 2, the delay tap and Doppler tap corresponding to the maximum value of $\hat h[k,l]$ are identified in the DD domain grid, denoted as the integer Doppler shift $k_{i,I}$ and integer delay $l_{i,I}$ of the $i$-th path. In step 5, based on the delay estimation step $\Delta \tau_s$ and Doppler estimation step $\Delta \upsilon_s$, the fractional Doppler shift $k_{i,F}$ and delay $l_{i,F}$ are iterated exhaustively within the range from $-$0.5 to 0.5, and the equivalent channel function ${h_{EQ,i}}\left[ {\Delta k,\Delta l} \right] = {h_{eq,{\upsilon _i}}}\left[ {\Delta k} \right] \cdot {h_{eq,{\tau _i}}}\left[ {\Delta l} \right]$ is established according to Eq. (17) for each Doppler index $k_{i,I} + k_{i,F}$ and delay index $l_{i,I} + l_{i,F}$. The two-dimensional correlation function between the measured CSF $\hat h[k,l]$ and the equivalent channel function ${h_{EQ,i}}\left[ {\Delta k,\Delta l} \right]$ is calculated as [46]

\begin{equation}\label{2D-corr-func}
\begin{split}
  &{R_{{h_{EQ,i}},\hat h}}\left[ {{k_{i,F}},{l_{i,F}}} \right]  \\ 
  &= \frac{1}{{MN}}\sum\limits_{n = 0}^{N - 1} {\sum\limits_{m = 0}^{M - 1} {h_{EQ,i}^*\left[ {n - {k_{i,F}},m - {l_{i,F}}} \right] \hat h\left[ {n,m} \right]} }.
\end{split}
\end{equation}
In step 8, when the fractional Doppler shift and delay match the true values, the peak of the two-dimensional correlation function attains its maximum. The coordinates of this peak determine the Doppler shift and delay of the $i$-th path. Further, the Doppler shift ${\hat \upsilon _i}$, delay ${\hat \tau _i}$, path gain ${\hat h_i}$ and phase ${\hat \varphi _i}$ of the $i$-th path can be derived from the following equations
\begin{equation}\label{fraction DD index}
  \left\langle {{{\hat k}_{i,F}},{{\hat l}_{i,F}}} \right\rangle  = \mathop {\arg \max }\limits_{ - 0.5 \le {k_{i,F}},{l_{i,F}} \le 0.5}  \left\{ {\left. {\left| {{R_{{h_{EQ,i}},\hat h}}\left[ {{k_{i,F}},{l_{i,F}}} \right]} \right|} \right\}} \right.,
  \vspace{-0.2cm}
\end{equation}

\begin{equation}\label{Doppler_hat}
  {\hat \upsilon _i} = \left( {{k_{i,I}} + {{\hat k}_{i,F}}} \right) \cdot \Delta \upsilon,
  \vspace{-0.2cm}
\end{equation}

\vspace{-0.2cm}
\begin{equation}\label{Delay_hat}
  {\hat \tau _i} = \left( {{l_{i,I}} + {{\hat l}_{i,F}}} \right) \cdot \Delta \tau,
  \vspace{-0.2cm}
\end{equation}

\vspace{-0.2cm}
\begin{equation}\label{h_hat}
  {\hat h_i} = \max \left\{ {\left. {\left| {{R_{{h_{EQ,i}},\hat h}}\left[ {{k_{i,F}},{l_{i,F}}} \right]} \right|} \right\}} \right.,
  \vspace{-0.2cm}
\end{equation}

\vspace{-0.2cm}
\begin{equation}\label{phi_hat}
  {\hat \varphi _i} =  - 2\pi {\hat \upsilon _i}{\hat \tau _i}.
  \vspace{-0.2cm}
\end{equation}

Finally, the power ${\left| {{{\hat h}_i}} \right|^2}$ of the $i$-th path is compared with the threshold $P_{th}$. If the power is lower than the threshold, the algorithm terminates. Otherwise, the path is identified as a valid MPC. Subsequently, the equivalent channel function of the $i$-th path is reconstructed and subtracted from the measured CSF to eliminate its interference, and the process repeats for the next iteration.

Notably, the successive cancellation procedure may suffer from error propagation, since inaccurate estimation of a strong MPC can leave residual components that affect the extraction of subsequent weak MPCs. Although the proposed fractional delay and Doppler shift estimation improves the reconstruction accuracy of each MPC, it cannot completely eliminate this inherent effect. Error propagation has also been reported for the CLEAN algorithm, for which further refinement can be performed, such as using the CLEAN estimates to initialize the SAGE algorithm for iterative parameter refinement [47], [48]. Following this principle, after the initial extraction of all MPCs, each MPC can be iteratively refined. Specifically, to refine the parameters of the $i$-th MPC, the contributions of all other estimated MPCs are reconstructed using their equivalent channel functions and subtracted from the original measured CSF. The proposed algorithm is then reapplied to the residual CSF to update the parameters of the $i$-th MPC. This procedure can be iteratively applied to all extracted MPCs until the reconstruction error converges or a predefined maximum number of iterations is reached, thereby mitigating error propagation at the expense of increased computational complexity.

\vspace{-0.3cm}
\subsection{Performance Analysis}
In this subsection, the performance of the proposed DD domain channel measurement method is evaluated in terms of PAPR, synchronization performance, CSF estimation performance, and fractional delay and Doppler shift estimation performance.
\begin{figure}[!t]
\center
\includegraphics[scale=0.4]{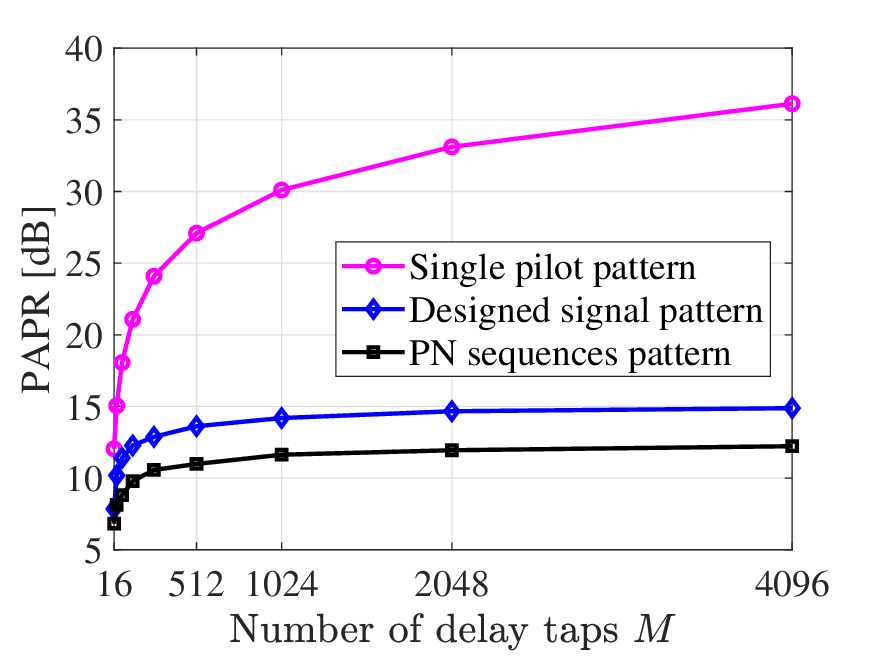}\\
\setlength{\abovecaptionskip}{-2pt}
\caption{PAPR of different signal patterns. \label{fig:2}}
\vspace{-0.45cm}
\end{figure}
\begin{figure}[!t]
\center
\includegraphics[scale=0.4]{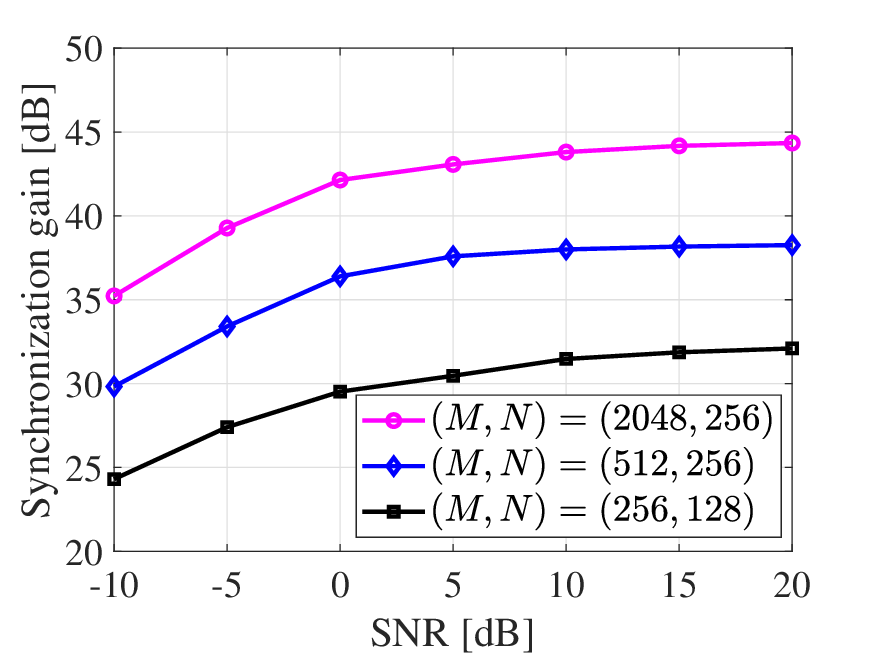}\\
\setlength{\abovecaptionskip}{-2pt}
\caption{Synchronization gains for various SNRs and OTFS frame sizes. \label{fig:3}}
\vspace{-0.45cm}
\end{figure}

\begin{figure}[!t]
{
\centering
\subfloat[]{
\includegraphics[scale=0.38]{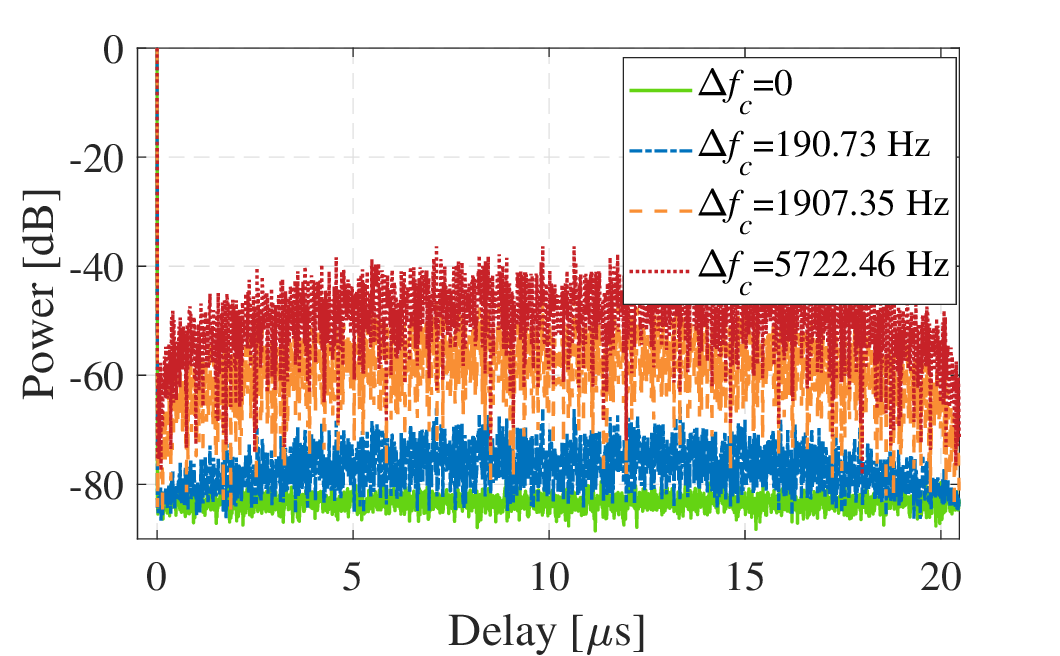}
}
\quad
\subfloat[]{
\includegraphics[scale=0.38]{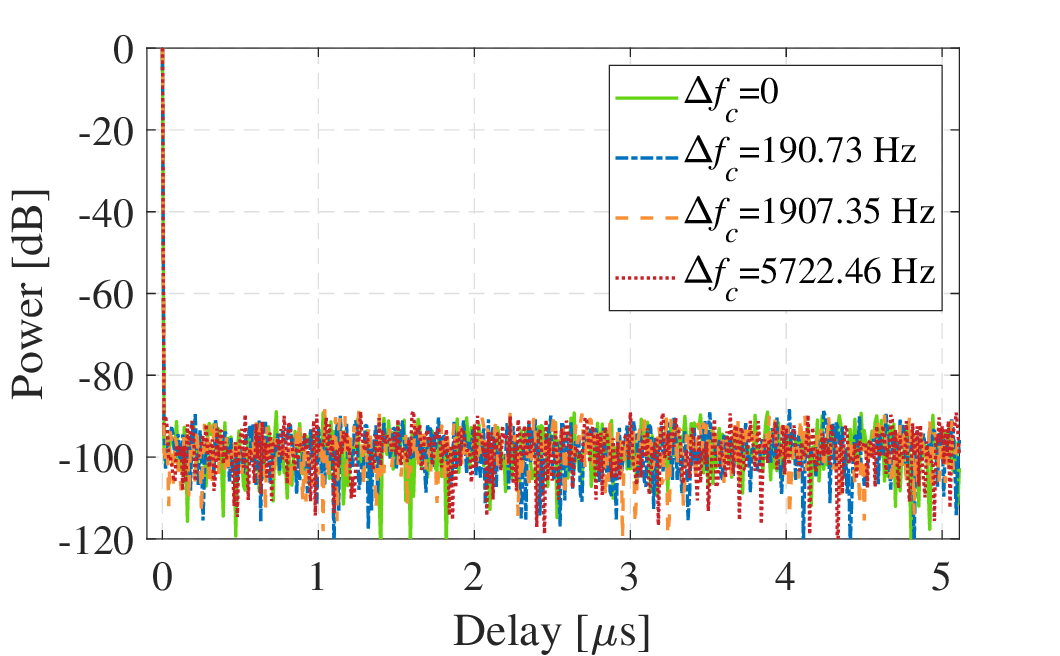}
}
\caption{Dynamic range of different channel measurement methods. (a) OFDM-based frequency domain method. (b) OTFS-based DD domain method. \label{fig:4}}
\vspace{-0.45cm}}
\end{figure}

\begin{figure}[!t]
{
\center
\includegraphics[scale=0.4]{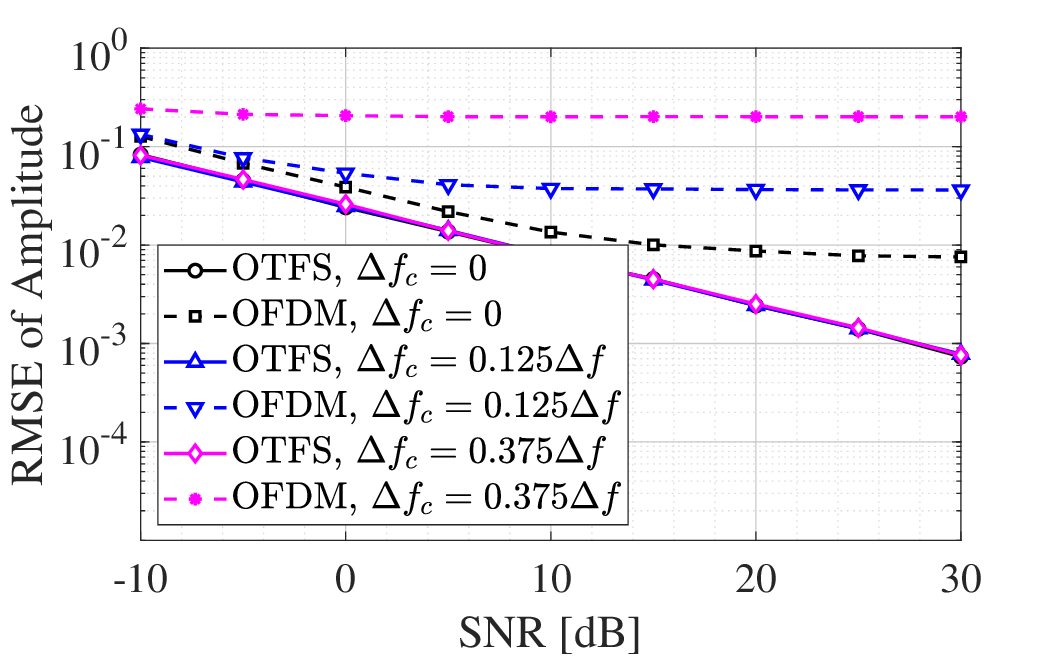}\\
\setlength{\abovecaptionskip}{-2pt}
\caption{RMSE of amplitude versus SNR under different levels of ICI. \label{fig:5}}
\vspace{-0.45cm}}
\end{figure}

\subsubsection{PAPR}
The PAPR of the sounding signal significantly influences the transmission efficiency and reception quality. High PAPR signals cause nonlinear distortion and impose stringent linearity requirements on the hardware. As shown in Fig. 2, the PAPRs of the sounding signal with three different DD domain patterns are compared, including the designed signal pattern, the single pilot pattern (containing a solitary pilot with guard symbols), and the PN sequences pattern (where the DD domain is fully occupied by PN sequences). In the simulation, the number of delay bins $M$ is set as a power of 2, ranging from 16 to 4096, with $N=M/2$. It can be observed that the PAPR increases with the number of delay bins and eventually stabilizes. The PAPR of the designed signal consistently remains below 15 dB, which is merely 3 dB higher than that of the PN sequences pattern, yet approximately 20 dB lower than that of the single pilot pattern. This indicates that the designed signal, owing to the insertion of the PN sequences, can effectively reduce the PAPR.

\subsubsection{Synchronization Performance}
To evaluate the performance of the proposed synchronization method, we adopt synchronization gain as the performance metric, which is defined as the ratio of the correlation peak to the noise level in an additive white Gaussian noise (AWGN) channel [49].
Fig. 3 shows the synchronization gain versus signal-to-noise ratio (SNR) for various OTFS frame sizes. It is evident that larger frame sizes yield higher synchronization gain. Furthermore, although the synchronization gain naturally decreases as SNR drops, reliable synchronization performance is maintained even in low SNR regimes. {In addition, a longer synchronization signal can provide a higher synchronization gain.}

{
The synchronization method based on sliding correlation may experience performance degradation at low SNR and under strong multipath and non line-of-sight (NLOS) propagation, where a later-arriving strong MPC may produce a larger correlation peak than the earliest MPC. In contrast, the method exhibits relatively strong robustness to carrier frequency offset (CFO) and an initial mismatch between the TX and RX sampling instants, referred to as sampling phase offset (SPO). Therefore, in practical measurements, the continuous transmission of the sounding signal allows synchronization to be performed under favorable propagation conditions before subsequent channel measurements.}

\begin{figure}[!t]
\centering
\subfloat[]{
\includegraphics[scale=0.38]{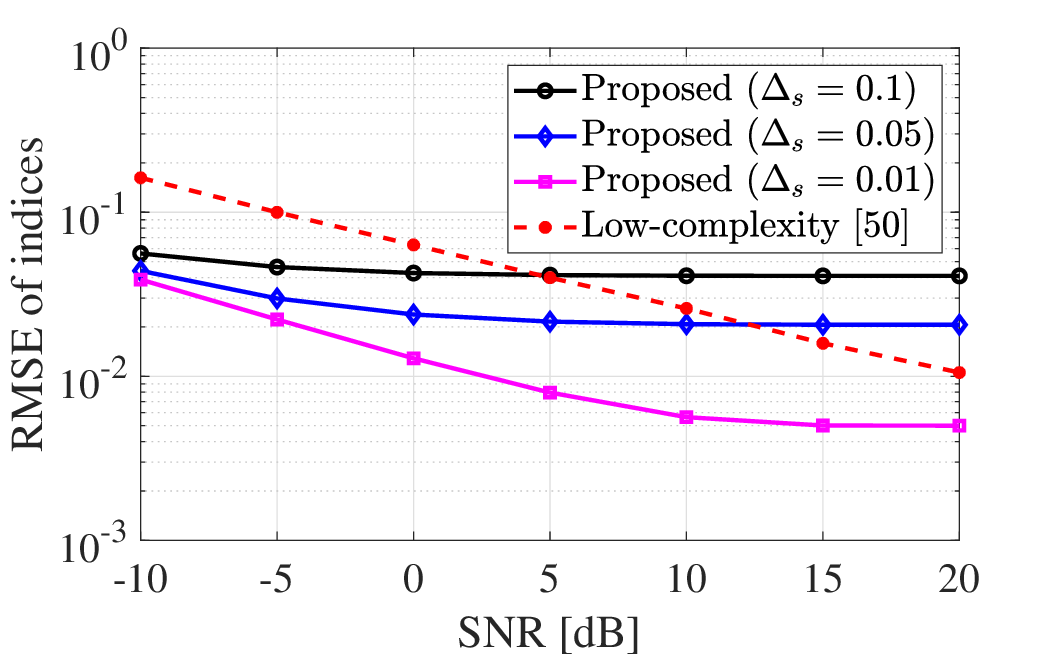}
}
\quad
\subfloat[]{
\includegraphics[scale=0.38]{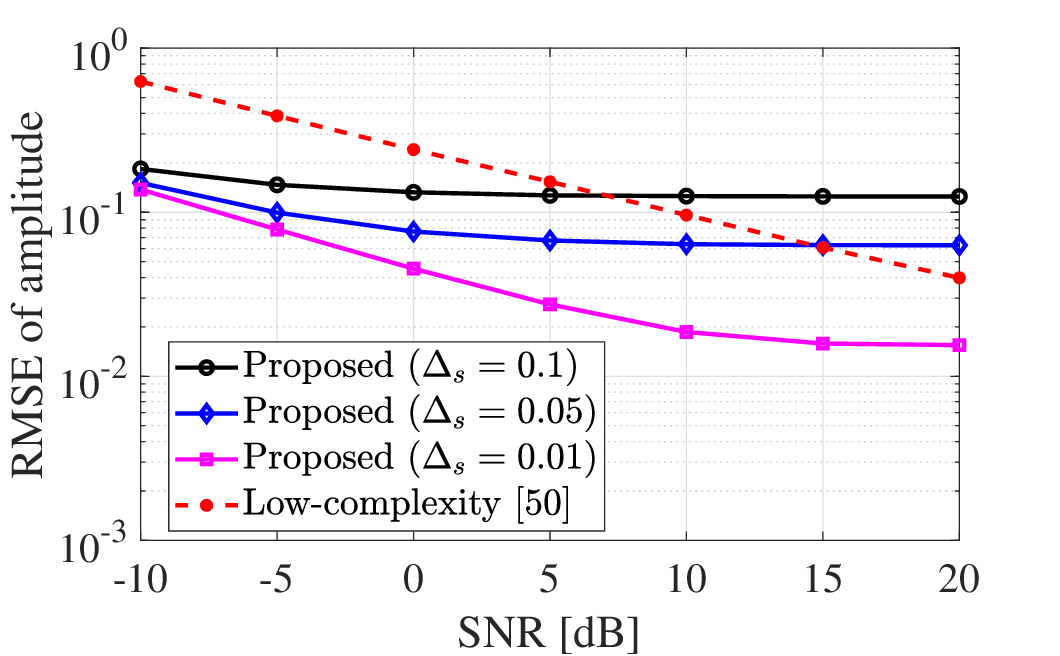}
}
\caption{RMSE performance of the joint fractional delay and Doppler shift estimation algorithm. (a) Estimation of delay and Doppler shift indices. (b) Estimation of multipath amplitude. \label{fig:6}}
\vspace{-0.45cm}
\end{figure}



{
\subsubsection{{CSF Estimation Performance}}
The CSF estimation performance is first evaluated in terms of dynamic range, which refers to the range of MPC powers that can be detected by a channel measurement method. If the dynamic range is limited, weak MPCs may be submerged in the noise and cannot be reliably extracted. For frequency domain channel measurement, a large CFO $\Delta {f_c}$ or Doppler shift destroys the orthogonality of the subcarriers, thereby degrading the dynamic range. As a DD domain modulation scheme, OTFS can obtain full diversity gain in the time and frequency domains [27]. Consequently, the dynamic range of the DD domain measurement method is less sensitive to CFO and Doppler shift. For comparison, the OFDM waveform employs the same number of subcarriers and bandwidth as the proposed OTFS waveform. Specifically, the adopted OTFS waveform has a frame size of $(2048, 256)$ and a bandwidth of 100 MHz, corresponding to a delay resolution of 10 ns and a Doppler resolution of 190.73 Hz, while the OFDM waveform employs 2048 subcarriers over the same bandwidth. Different values of $\Delta {f_c}$, set as integer multiples of the Doppler resolution of the OTFS system, are evaluated in an ideal channel. As shown in Fig. 4, the dynamic range of the OFDM-based frequency domain channel measurement method decreases with increasing $\Delta {f_c}$, whereas that of the proposed DD domain method remains relatively stable and maintains a large dynamic range. For the proposed method, the influence of $\Delta {f_c}$ is mainly manifested as a shift of the channel response along the Doppler axis.}

{
In addition to the dynamic range, the MPC amplitude estimation accuracy is further evaluated by comparing the proposed method with the OFDM-based frequency domain channel measurement method under the same bandwidth and number of subcarriers. To exclude the influence of delay domain overlap, five delay-resolvable MPCs with delays of [0, 1, 2, 3, 4] $\mu s$ and relative powers of [0, $-$10, $-$15, $-$20, $-$25] dB are considered. Different CFO values of 0, 0.125$\Delta {f}$, and 0.375$\Delta {f}$ are introduced to generate different levels of ICI. The root mean square error (RMSE) of the estimated MPC amplitude is adopted as the performance metric. As shown in Fig. 5, the RMSE of the estimated MPC amplitude for the OFDM-based method increases as the ICI becomes more severe and exhibits an error floor at high SNR due to the loss of subcarrier orthogonality. In contrast, the proposed DD domain channel measurement method is much less affected by ICI, and the RMSE of the estimated MPC amplitude continuously decreases with increasing SNR. Therefore, the proposed method exhibits improved robustness to ICI in terms of MPC amplitude estimation compared with the OFDM-based frequency domain channel measurement method.
}


\subsubsection{Fractional Delay and Doppler Shift Estimation Performance}
To evaluate the performance of the joint fractional delay and Doppler shift estimation algorithm, the impact of noise and estimation steps is analyzed by varying the delay and Doppler shift of the channel. The performance is assessed from two aspects: the estimation error of the multipath delay and Doppler shift indices, and the estimation error of the multipath amplitude. The RMSE is adopted as the performance metric.

In the simulation, the fractional delay and Doppler shift traverse the range from $-$0.5 to 0.5 times the delay and Doppler resolutions. The delay estimation step $\Delta {\tau _{{s}}}$ and Doppler estimation step $\Delta {\upsilon _{{s}}}$ are set to an equal value $\Delta_{{s}}$, which takes values of 0.1, 0.05, and 0.01. 
Fig. 6 compares the proposed algorithm with the low-complexity algorithm in [50]. It can be observed that the proposed algorithm exhibits superior estimation accuracy compared to the low-complexity algorithm. Moreover, it maintains robust performance in the low SNR regime, even with a relatively large estimation step. 

{In addition to the above estimation performance, the computational complexity, runtime, and scalability of the proposed algorithm are further evaluated.
The computational complexity of the proposed joint fractional delay and Doppler shift estimation algorithm is dominated by the two-dimensional search over the fractional delay and Doppler shift, which is implemented through repeated evaluations of the two-dimensional correlation function. 
Let ${Q_\tau } = 1/\Delta {\tau _s}$ and ${Q_\upsilon } = 1/\Delta {\upsilon _s}$ denote the numbers of fractional delay and Doppler candidates, respectively. For each MPC, ${Q_\tau }{Q_\upsilon }$ candidate fractional delay and Doppler pairs are evaluated, and each correlation calculation over the $M \times N$ CSF has a complexity of $O\left( {MN} \right)$. Therefore, the dominant computational complexity for extracting $P$ MPCs is $O\left( {PMN{Q_\tau }{Q_\upsilon }} \right)$, which scales linearly with $P$ for fixed DD grid size and estimation steps. The successive path cancellation is performed after each MPC extraction and does not change the dominant complexity order. The runtime was evaluated on a computer equipped with an Intel Core i5-13500H CPU and 16.0 GB RAM. With $M = 2048$, $N = 256$, ${Q_\tau } = 10$, and ${Q_\upsilon } = 100$, the average runtime increased from 7.78 s for 5 MPCs to 53.54 s for 35 MPCs, and linear fitting of the measured average runtimes yielded $T\left( P \right) = 1.50P + 0.22$s with ${R^2} = 0.9991$. The results are consistent with the theoretical linear dependence on the number of MPCs and demonstrate approximately linear computational scalability with respect to $P$. Overall, the performance evaluation indicates that the proposed algorithm is well-suited for refined channel parameter estimation during the offline processing of DD domain channel measurements.
}

\begin{figure}[!t]
\center
\includegraphics[scale=0.28]{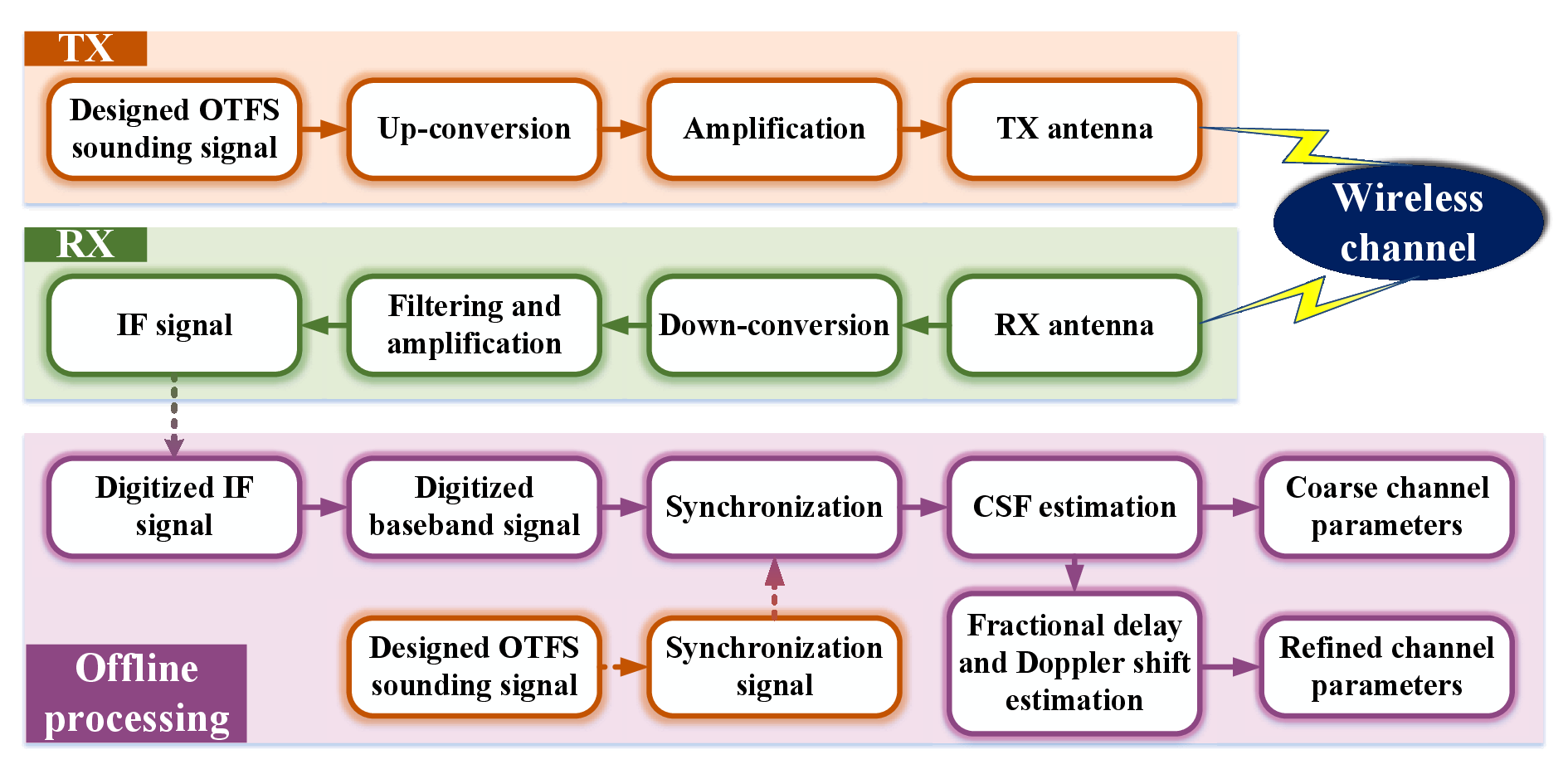}\\
\setlength{\abovecaptionskip}{-2pt}
\caption{The block diagram of the DD domain measurement system. \label{fig:7}}
\vspace{-0.45cm}
\end{figure}

\begin{figure}[!t]
\center
\includegraphics[scale=0.5]{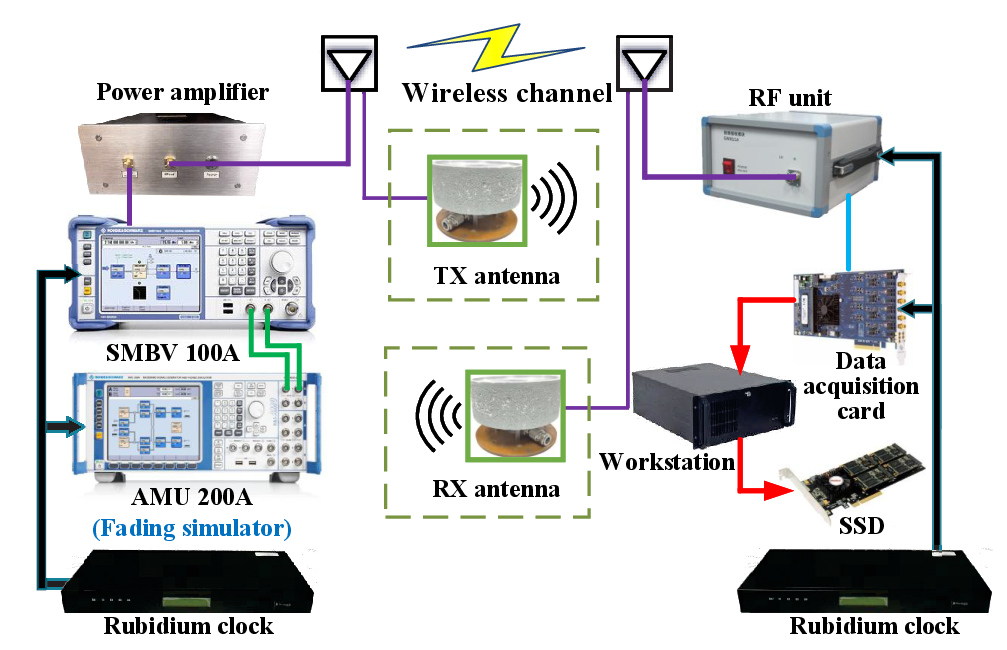}\\
\setlength{\abovecaptionskip}{-2pt}
\caption{The hardware platform of the DD domain measurement system. \label{fig:8}}
\vspace{-0.45cm}
\end{figure}

\section{DD Domain Channel Measurement System}
This section details the setup of the DD domain channel measurement system and subsequently presents the system calibration and verification results.

\subsection{System Setup}
Based on the proposed methodology, a DD domain channel measurement system is established. Fig. 7 presents the block diagram of the developed system. At the TX, the designed OTFS sounding signal is loaded into the radio frequency (RF) signal generator, up-converted, and adjusted for transmission power before being radiated through the TX antenna into the wireless channel. At the RX, the sounding signal is captured by the RX antenna and processed by the RF unit via down-conversion, filtering, and amplification to obtain the intermediate frequency (IF) signal. The IF signal is then sampled by a high-speed data acquisition card. During the offline processing stage, the digitized IF signal is processed to extract the baseband signal. A locally stored synchronization signal, derived from the designed sounding signal, is then utilized to achieve synchronization. Subsequently, the CSF is estimated and coarse channel parameters are obtained. By applying the proposed joint fractional delay and Doppler shift estimation algorithm to the CSF, refined channel parameters are extracted, thereby enabling accurate DD domain channel measurements.

Fig. 8 shows the hardware platform of the proposed DD domain channel measurement system. The TX consists of a baseband signal generator (AMU 200A), an RF signal generator (SMBV 100A), a power amplifier, an omnidirectional antenna, and a rubidium clock. Notably, the AMU 200A integrates a fading module capable of emulating typical time-varying multipath fading channels, which is primarily utilized for system verification. The RX consists of an omnidirectional antenna, an RF unit, a data acquisition card, a solid-state disk (SSD), a workstation, and a rubidium clock. The RX antenna is connected to the RF unit, which outputs a 180 MHz IF signal. Automatic gain control of the RF unit can be disabled to allow for manual RF gain configuration. The output of the RF unit is transferred to the high-speed data acquisition card, which is controlled by the workstation. The sampling rate of the data acquisition card can be set to 250 million samples per second (Msps) or 500 Msps to ensure that the IF signal does not have spectrum aliasing. Finally, the sampled digital IF signals are stored in the high-speed SSD of the workstation, and subsequently demodulated, low-pass filtered, and resampled to obtain the baseband received signal. Rubidium clocks at both TX and RX ensure frequency synchronization across the entire system.

\begin{figure}[!t]
\centering
\subfloat[]{
\includegraphics[scale=0.36]{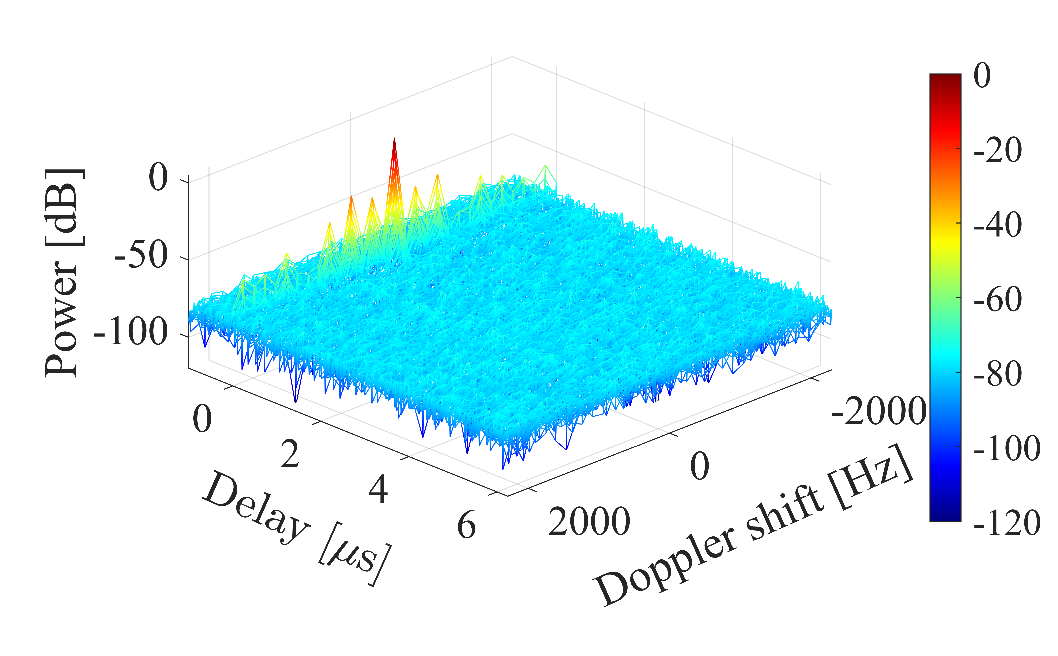}
}
\quad
\subfloat[]{
\includegraphics[scale=0.36]{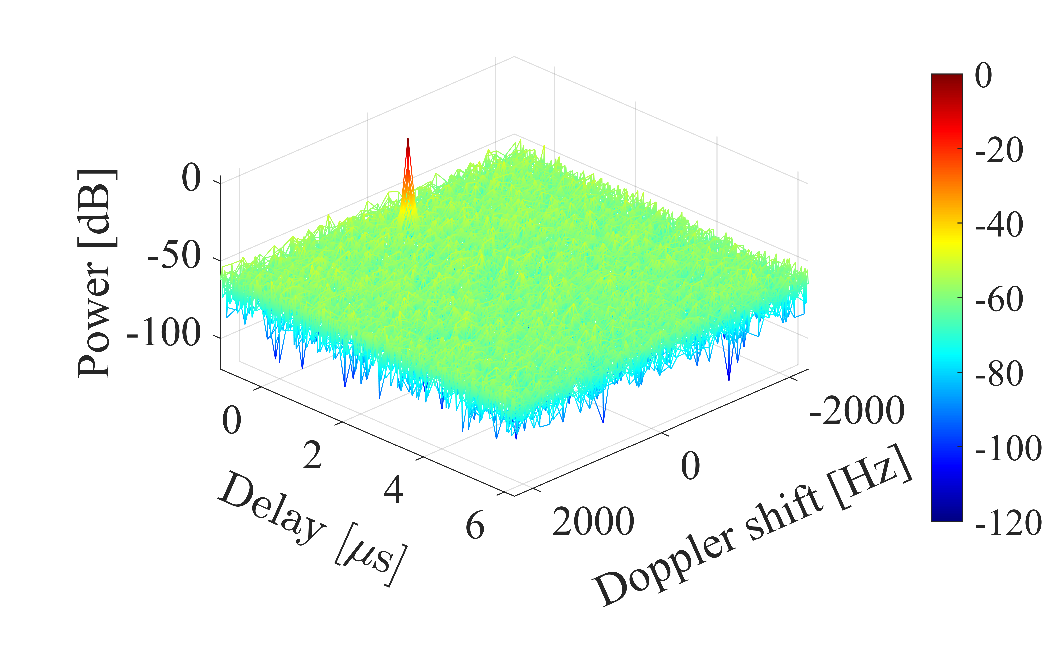}
}
\caption{B2B calibration. (a) Before calibration. (b) After calibration. \label{fig:8}}
\vspace{-0.45cm}
\end{figure}

\begin{figure}[!t]
\centering
\includegraphics[scale=0.35]{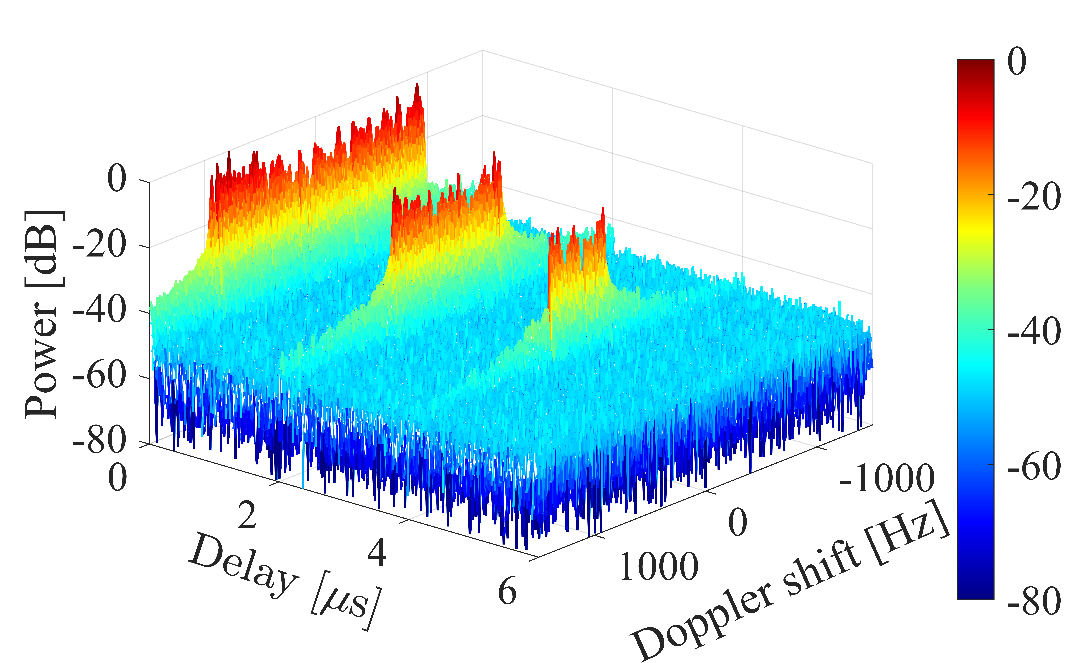}
\caption{Verification results under the Rayleigh channel. \label{fig:9}}
\vspace{-0.45cm}
\end{figure}

\subsection{System Calibration and Verification}
To ensure the reliability of DD domain channel measurements, preliminary calibrations and verifications are required prior to field measurement campaigns. A back-to-back (B2B) calibration is conducted in the absence of antennas, where the RF signal generator is directly connected to the RF unit via cables. Through this setup, the inherent system response introduced by the equipment, cables, and RF interfaces is acquired and subsequently eliminated in the time-frequency domain. Fig. 9 illustrates the B2B calibration results. As observed, the spurious multipath components introduced by the hardware are effectively eliminated after calibration. Consequently, a more accurate DD domain CSF can be obtained during actual measurements.

{Although the B2B calibration compensates for the deterministic system response introduced by the equipment, cables, and RF interfaces, several residual hardware impairments may still affect the measurement accuracy. Phase noise may degrade phase coherence and broaden the measured response in the Doppler domain, thereby affecting Doppler shift and amplitude estimation. Clock drift may introduce time-varying timing offsets and affect MPC delay estimation [51]. In the developed system, high-precision rubidium clocks provide external frequency references to the TX and RX hardware and corresponding sampling clocks, improving frequency and timing stability. RF nonlinearity may introduce amplitude distortion and spurious components, and the transmission power and receiver gain are configured to maintain the RF chain within its linear operating range. The antenna pattern may introduce direction-dependent gain variations and consequently affect the measured MPC amplitudes. In the current measurements, omnidirectional antennas are employed to reduce such variations, while additional antenna-pattern characterization and calibration would be required for future directional or multi-antenna measurements [52]. Gain variations are mitigated by disabling the automatic gain control and using a fixed RF gain. Limited analog-to-digital converter (ADC) quantization resolution may reduce the effective dynamic range and affect weak-MPC extraction, while a 14-bit ADC is employed in the developed system.}

Following the calibration, the system verification is conducted within the identical B2B setup by activating the built-in fading module of the baseband signal generator. To comprehensively evaluate the proposed method, the verification results under a Rayleigh channel and a pure Doppler shift channel are presented below.

\begin{figure}[!t]
{
\centering
\subfloat[]{
\includegraphics[scale=0.26]{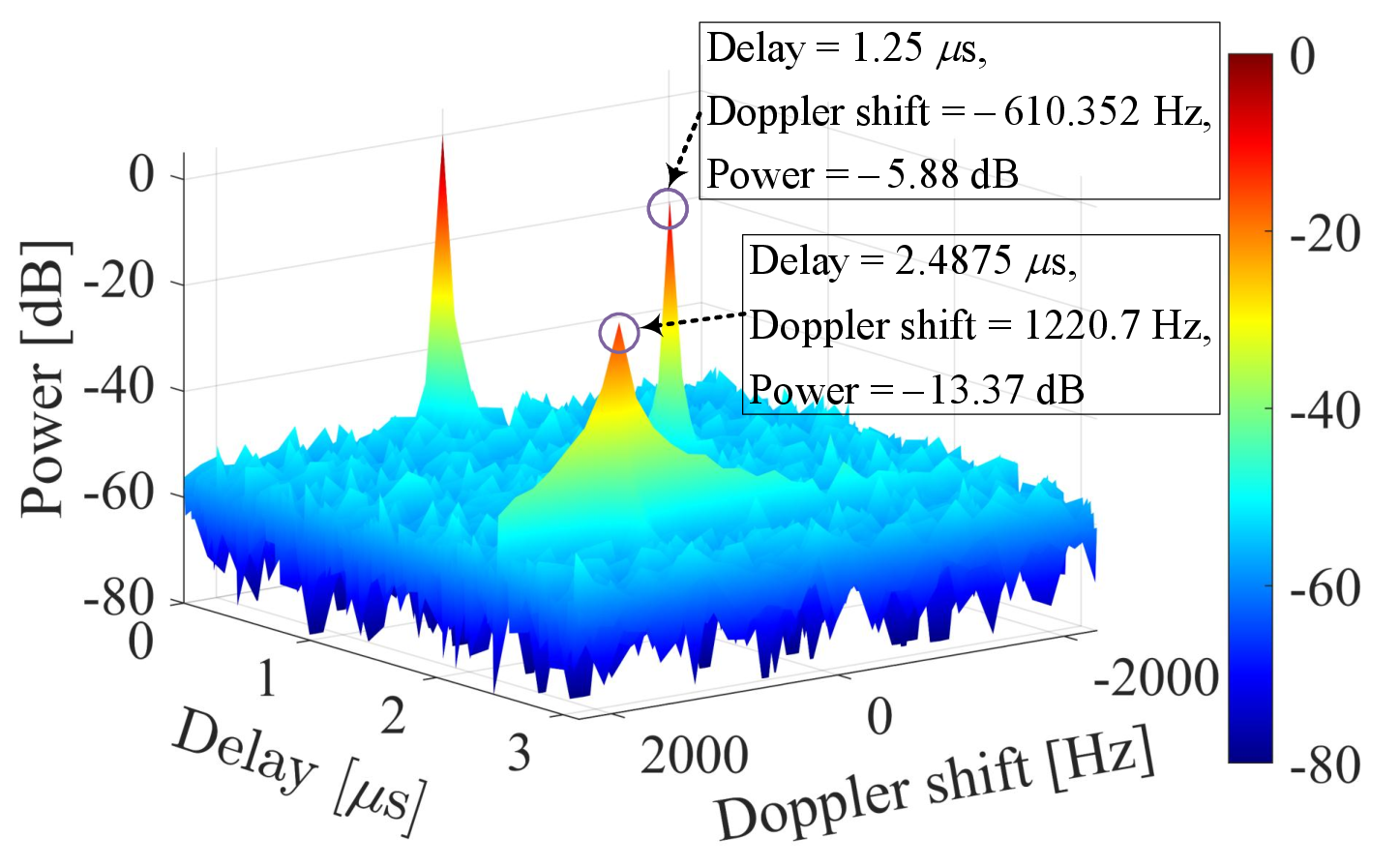}
}
\quad
\subfloat[]{
\includegraphics[scale=0.26]{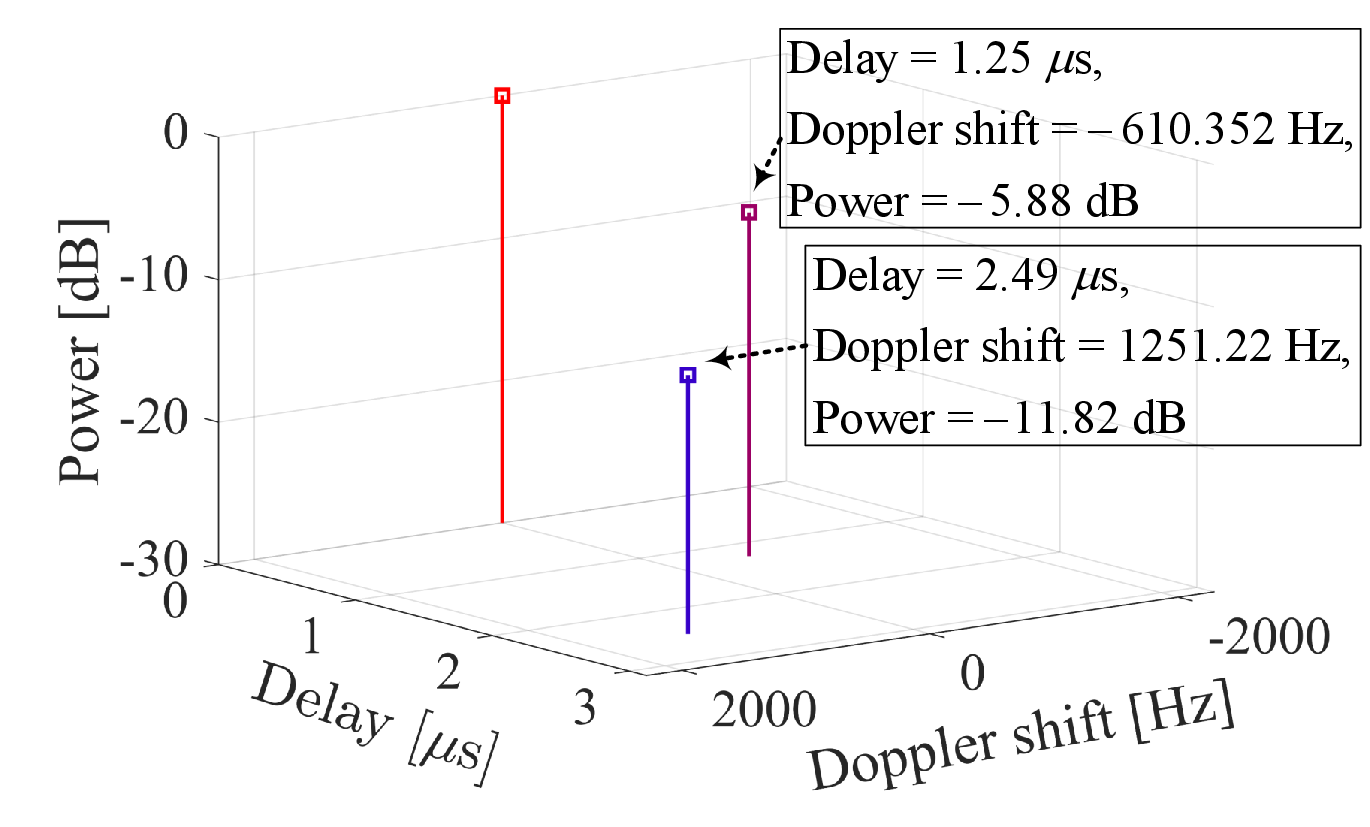}
}
\caption{Verification results under the pure Doppler shift channel. (a) CSF. (b) Fractional delay and Doppler shift estimation. \label{fig:11}}
\vspace{-0.45cm}}
\end{figure}

\subsubsection{Rayleigh Channel}
To verify the proposed channel measurement method, we focus on the measured CSF. The fading module is configured to simulate a Rayleigh channel with three distinct Jakes spectra. The path delays are set to 0, 2, and 4 $\mu s $, with corresponding powers of 0, $-$5, and $-$10 dB, respectively. To avoid fractional Doppler effects, the maximum Doppler shifts for each Jakes spectrum are set as integer multiples of the Doppler resolution, specifically 953.67, 476.84, and 238.42 Hz. In this setup, the adopted sounding signal has a frame size of $(4096, 2048)$ and a bandwidth of 100 MHz, {corresponding to a delay resolution of 10 ns and a Doppler resolution of 11.92 Hz. Accordingly, the preset path delays of 0, 2, and 4 $\mu s$ correspond to [0, 200, 400]$\Delta \tau$, while the maximum Doppler shifts of 953.67, 476.84, and 238.42 Hz correspond to [80, 40, 20]$\Delta \upsilon$, respectively}. As can be seen from Fig. 10, the delay, maximum Doppler shift, and power of each path closely match the preset configurations, which confirms that the proposed method ensures accurate measurements.

\subsubsection{Pure Doppler Shift Channel}
To further verify the joint fractional delay and Doppler shift estimation algorithm, we concentrate on the parameters of the MPCs. The fading module is configured to simulate a pure Doppler shift channel with three independent paths. The path delays are set to 0, 1.25, and 2.49 $\mu s $, the Doppler shifts are set to 0, $-$610.35, and 1251.22 Hz, and the powers are set to 0, $-$5, and $-$10 dB, respectively. In this setup, the adopted sounding signal has a frame size of $(2048, 256)$ and a bandwidth of 80 MHz, {corresponding to a delay resolution of 12.5 ns and a Doppler resolution of 152.59 Hz. Accordingly, the preset path delays of 0, 1.25, and 2.49 $\mu s$ correspond to [0, 100, 199.2]$\Delta \tau$, while the Doppler shifts of 0, $-$610.35, and 1251.22 Hz correspond to [0, $-$4, 8.2]$\Delta \upsilon$, respectively}. Fig. 11(a) shows the measured CSF of the pure Doppler shift channel, where three distinct peaks corresponding to the configured paths are clearly visible. Notably, due to the fractional delay and Doppler shift present in the third path, power leakage occurs across both the delay and Doppler dimensions. Consequently, the measured power of the third path decreases to $-$13.37 dB, which is significantly lower than the preset value of $-$10 dB. As depicted in Fig. 11(b), the application of the proposed joint fractional delay and Doppler shift estimation algorithm enables the parameters of all MPCs to be accurately extracted, which confirms that the proposed DD domain channel measurement method is highly effective for channel measurement campaigns in high-mobility scenarios.

\section{{DD Domain Channel Measurements in V2X Scenarios}}
\subsection{{V2I Channel Measurements}}
\subsubsection{{Measurement Campaigns}}
The channel measurements are conducted at Gaoliangqiao Xiejie, Haidian District, Beijing, focusing on a V2I scenario. To avoid interference from commercial 5G frequency bands operated by China Mobile and China Unicom, the measurements are carried out in the 3.3-3.4 GHz band. The sounding signal employs a frame size of $\left( {2048,{\rm{ }}256} \right)$, and the transmission power is set to 28.38 dBm. The measurement equipment is illustrated in Fig. 12. During the measurements, the RX devices and RX antenna are positioned on an overpass at a height of 5 m. Meanwhile, the TX devices are installed in a private car, with the TX antenna mounted on its roof.

\begin{figure}[!t]
{
\center
\includegraphics[scale=0.24]{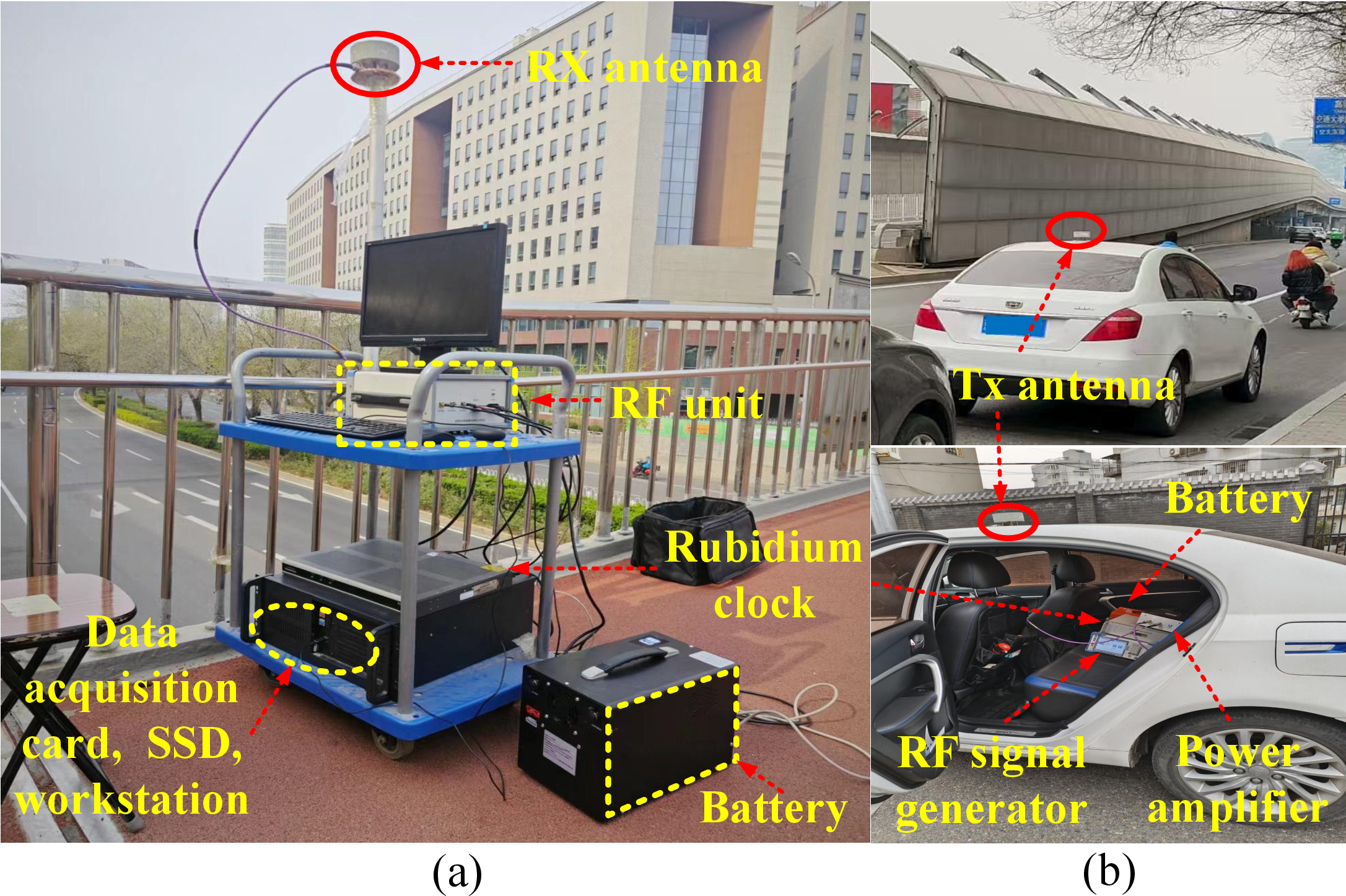}\\
\setlength{\abovecaptionskip}{-2pt}
\caption{Measurement equipment. (a) RX. (b) TX. \label{fig:12}}
\vspace{-0.45cm}}
\end{figure}

\begin{figure*}[!t]
{
\center
\includegraphics[scale=0.24]{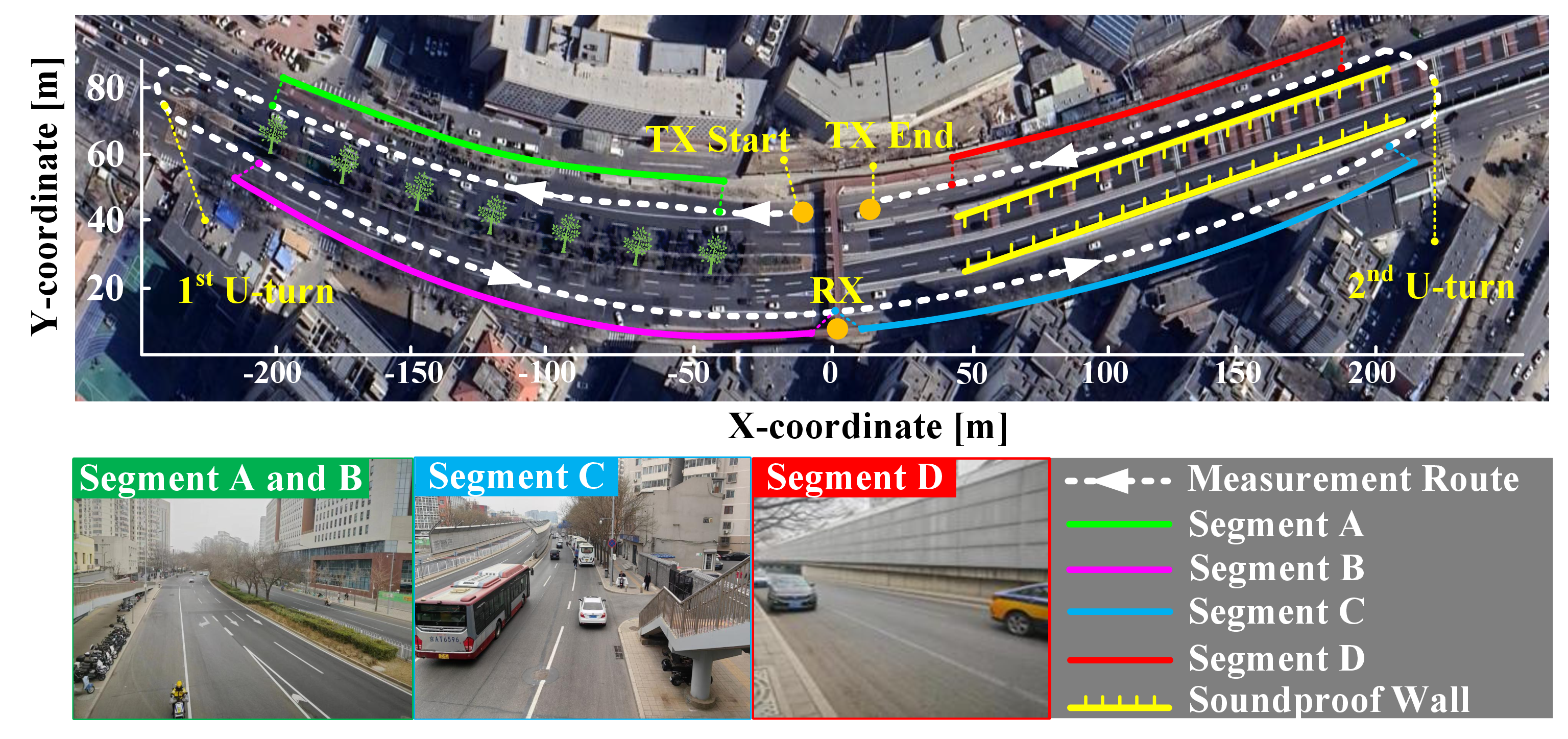}\\
\setlength{\abovecaptionskip}{-2pt}
\caption{Measurement route and environment. \label{fig:13}}
\vspace{-0.45cm}}
\end{figure*}

The measurement route and environment are illustrated in Fig. 13. The TX vehicle originates from a designated roadside location, executes two U-turns along the trajectory, and eventually returns to its starting point. To facilitate a comprehensive analysis, the measurement route is logically partitioned into four distinct segments according to the actual distance between the TX and RX (T-R distance) and propagation conditions. A more detailed description of the four segments is as follows.

\setlength{\parskip}{0pt}
\textbf{Segment A}: As the TX sets off from the starting position, the sounding signal is significantly influenced by the vegetation in the middle of the road, resulting in obstructed line-of-sight (OLOS) propagation. {During this segment, the TX reaches a maximum speed of approximately 50 km/h.} In this segment, the T-R distance ranges from $-$40 m to $-$250 m, where the negative sign indicates the relative spatial direction, and the absolute value represents the actual T-R distance.

\textbf{Segment B}: This segment corresponds to the route from the first U-turn until the TX approaches the RX, with the T-R distance varying from $-$250 m to 0 m. {The TX travels at a maximum speed of approximately 55 km/h along this route.} Due to the road curvature, some obstructions persist when the TX is at a large distance from the RX. As the TX continuously approaches the RX, line-of-sight (LOS) propagation becomes dominant.

\textbf{Segment C}: This segment represents the portion where the TX moves away from the position directly below the RX, prior to executing the second U-turn, {during which the maximum vehicle speed is approximately 58 km/h}. In this segment, the T-R distance increases from 0 m to 250 m, which is characterized by highly favorable LOS propagation.

\textbf{Segment D}: This segment covers the trajectory after the TX executes the second U-turn until it reaches the ending position, {with the TX reaching a maximum speed of approximately 48 km/h}. Due to the presence of the soundproof wall, the propagation conditions in this segment are NLOS, and the T-R distance decreases from 240 m to 48 m.

\subsubsection{{Measurement Results}}
In this section, we present the measurement results, including the CSF, PDP, and DPSD, as well as the number of MPCs, KF, RMS DS, and RMS DPS.

\begin{figure}[!t]
{
\centering
\subfloat[]{
\includegraphics[scale=0.26]{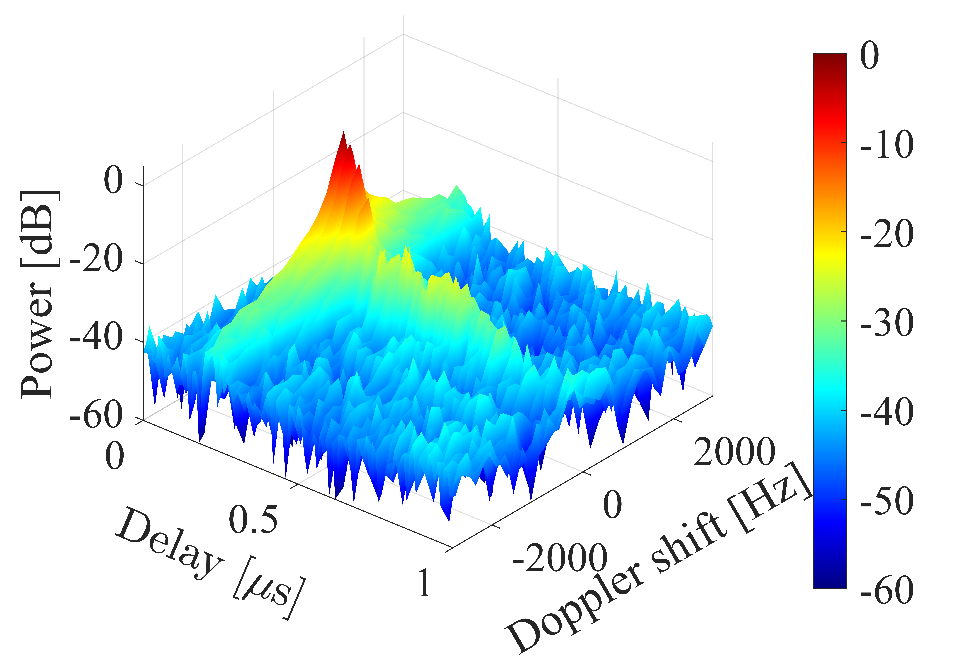}
}
\subfloat[]{
\includegraphics[scale=0.26]{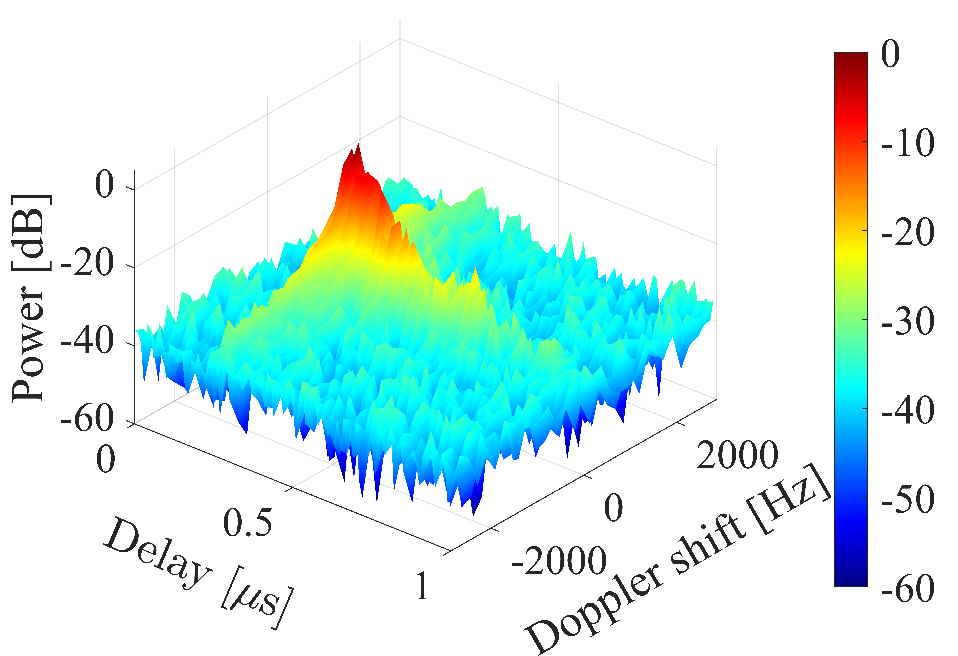}
}
\quad
\subfloat[]{
\includegraphics[scale=0.29]{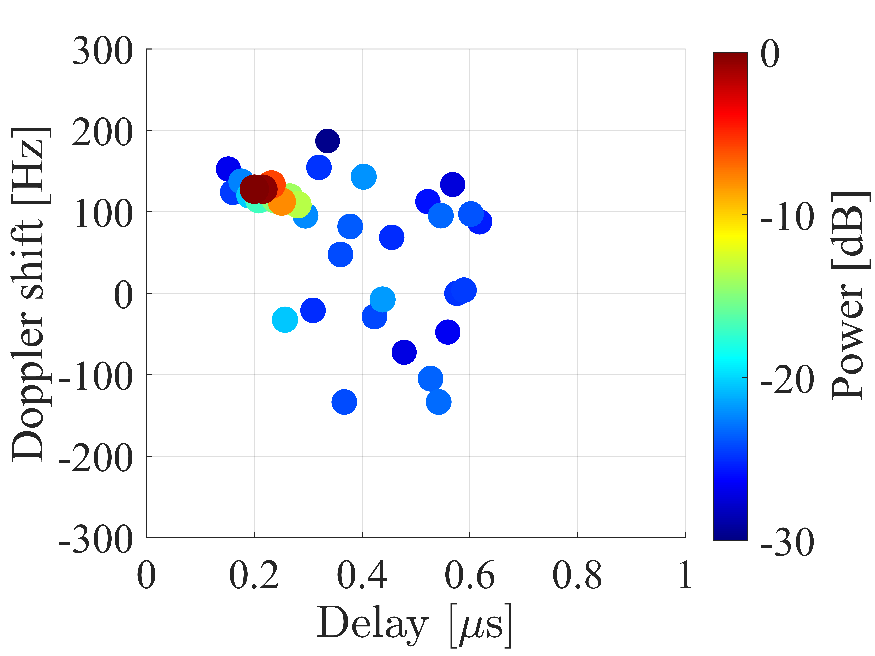}
}
\subfloat[]{
\includegraphics[scale=0.29]{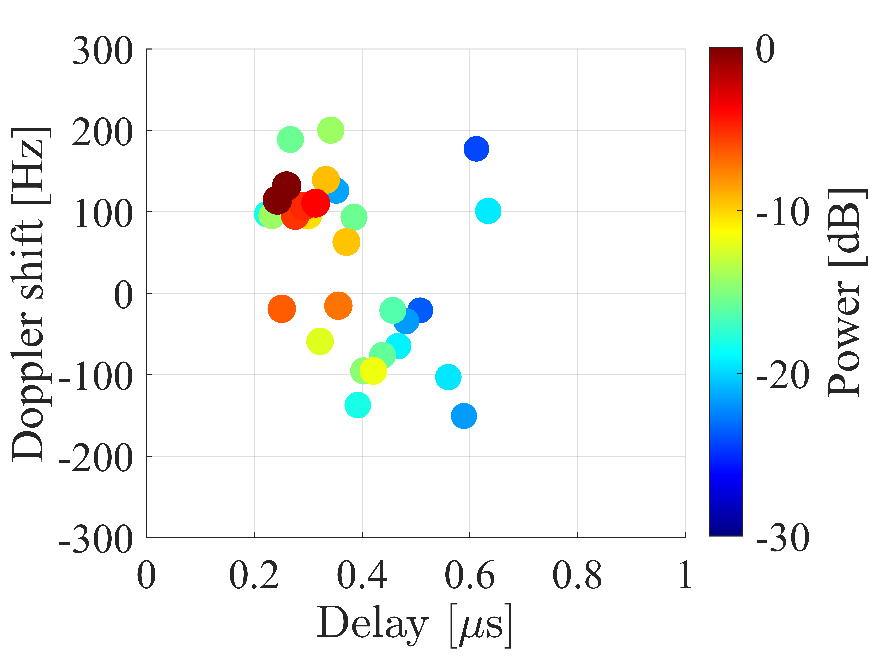}
}
\caption{Examples of CSFs and fractional delay and Doppler shift estimation results. (a) CSF for the LOS case. (b) CSF for the NLOS case. (c) Estimation results for the LOS case. (d) Estimation results for the NLOS case. \label{fig:14}}
\vspace{-0.45cm}}
\end{figure}

\begin{figure*}[!t]
\centering
\subfloat[]{
\includegraphics[scale=0.33]{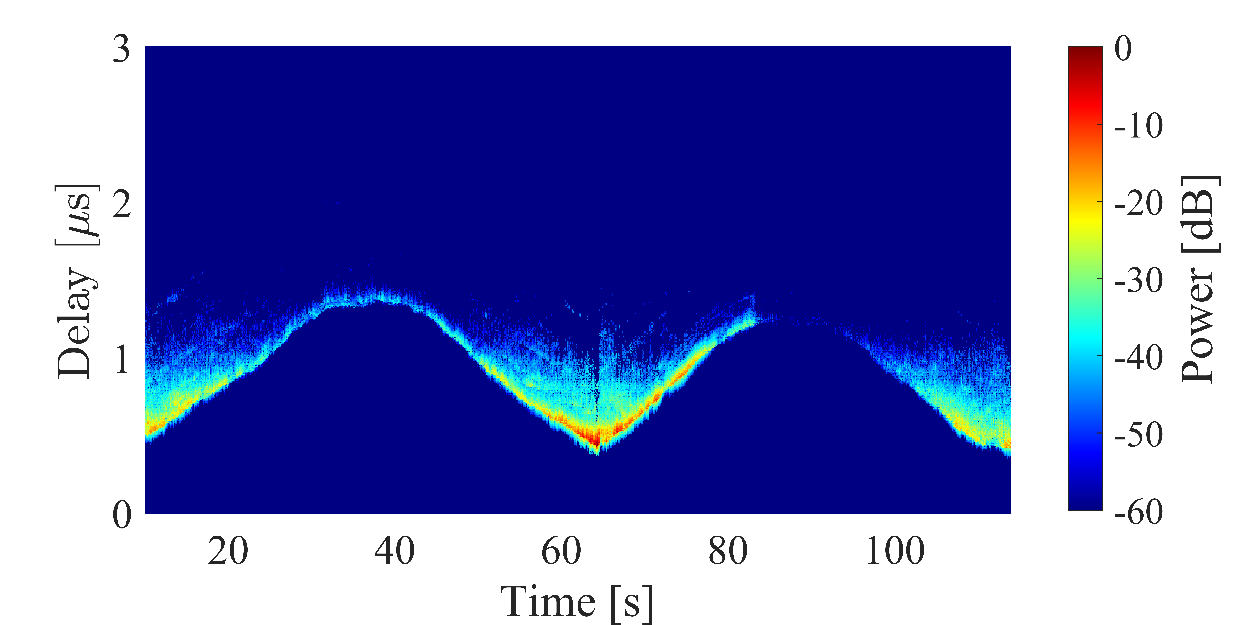}
}
\subfloat[]{
\includegraphics[scale=0.33]{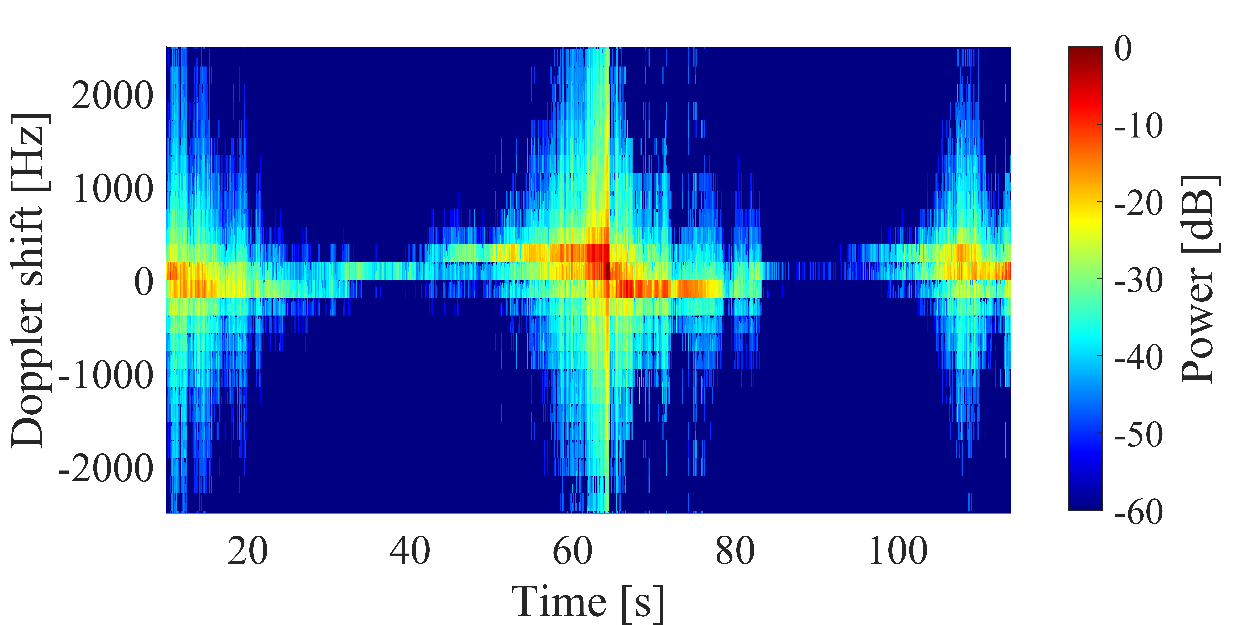}
}
\quad
\vskip -10pt
\subfloat[]{
\includegraphics[scale=0.33]{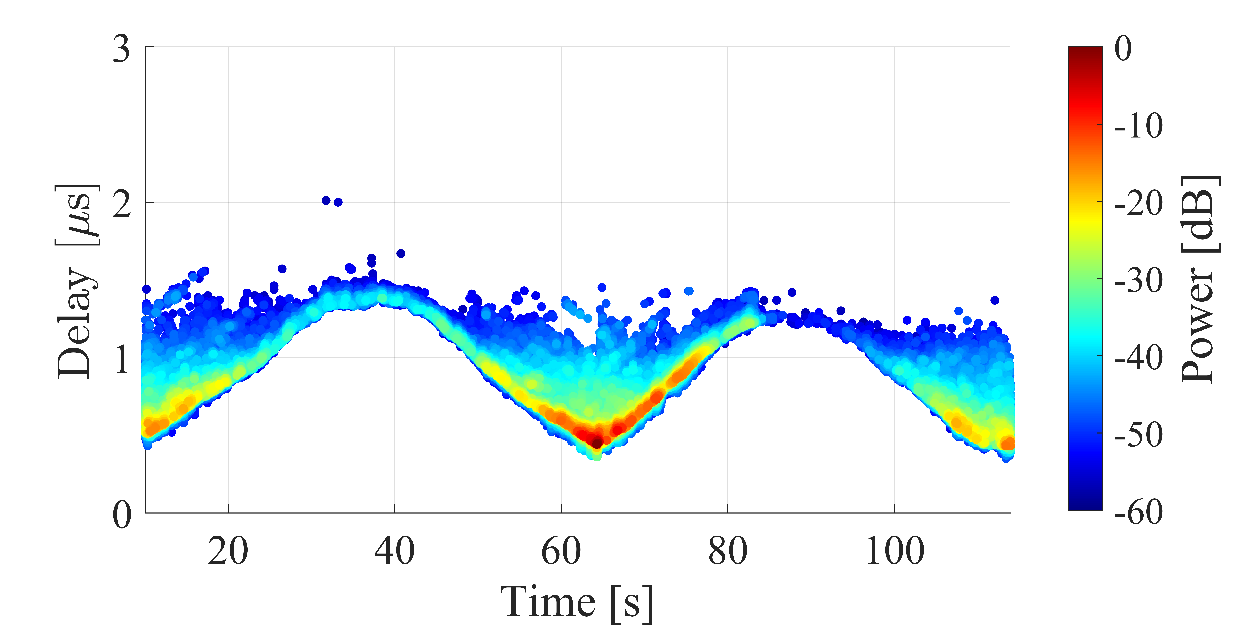}
}
\subfloat[]{
\includegraphics[scale=0.33]{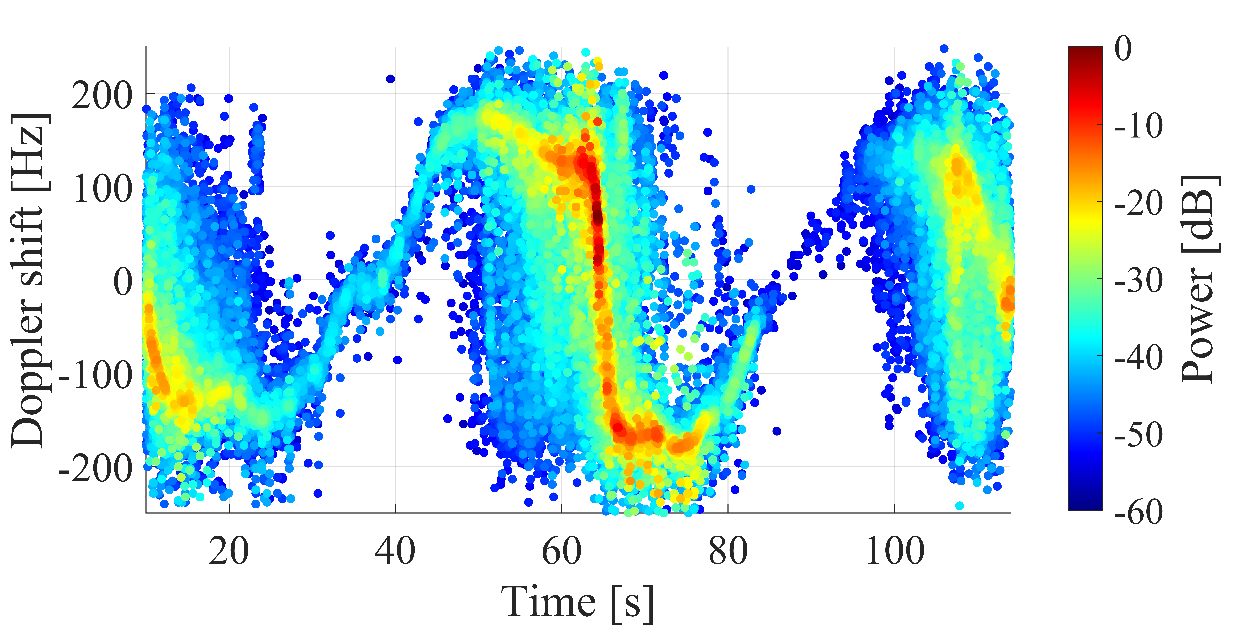}
}
\caption{Measured time-variant PDPs and DPSDs over the entire measurement. (a) PDP and (b) DPSD without fractional delay and Doppler shift estimation. (c) PDP and (d) DPSD with fractional delay and Doppler shift estimation. \label{fig:15}}
\vspace{-0.3cm}
\end{figure*}

\begin{figure*}[!t]
\centering
\subfloat[]{
\includegraphics[scale=0.32]{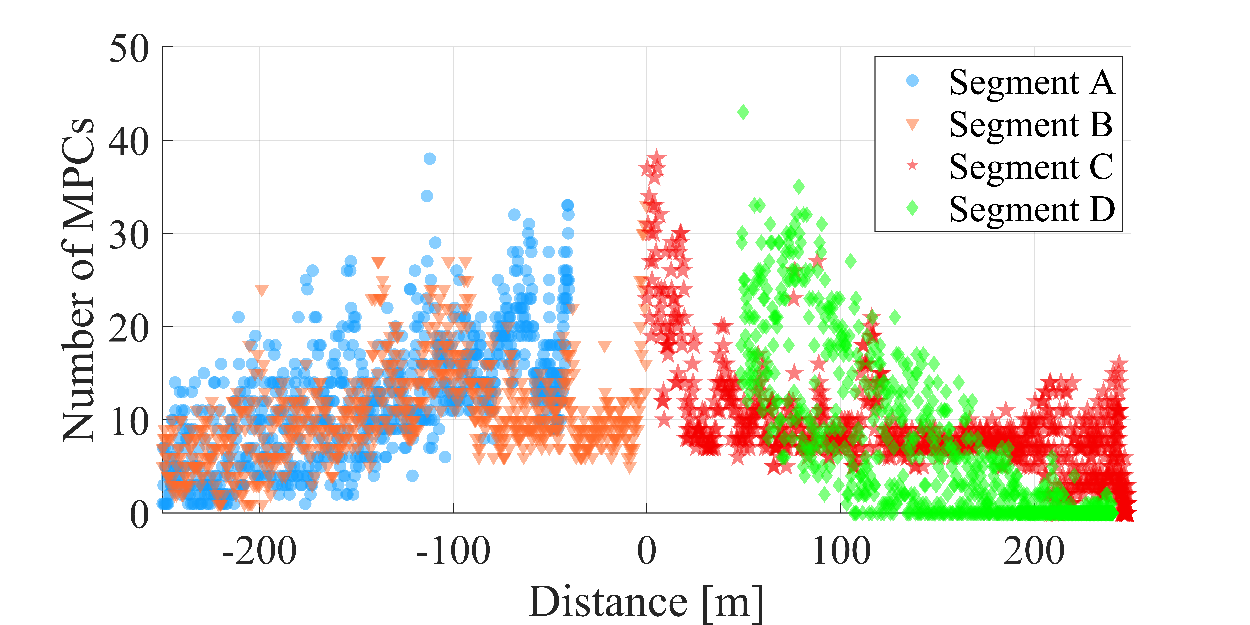}
}
\hspace{-0.2cm}
\subfloat[]{
\includegraphics[scale=0.32]{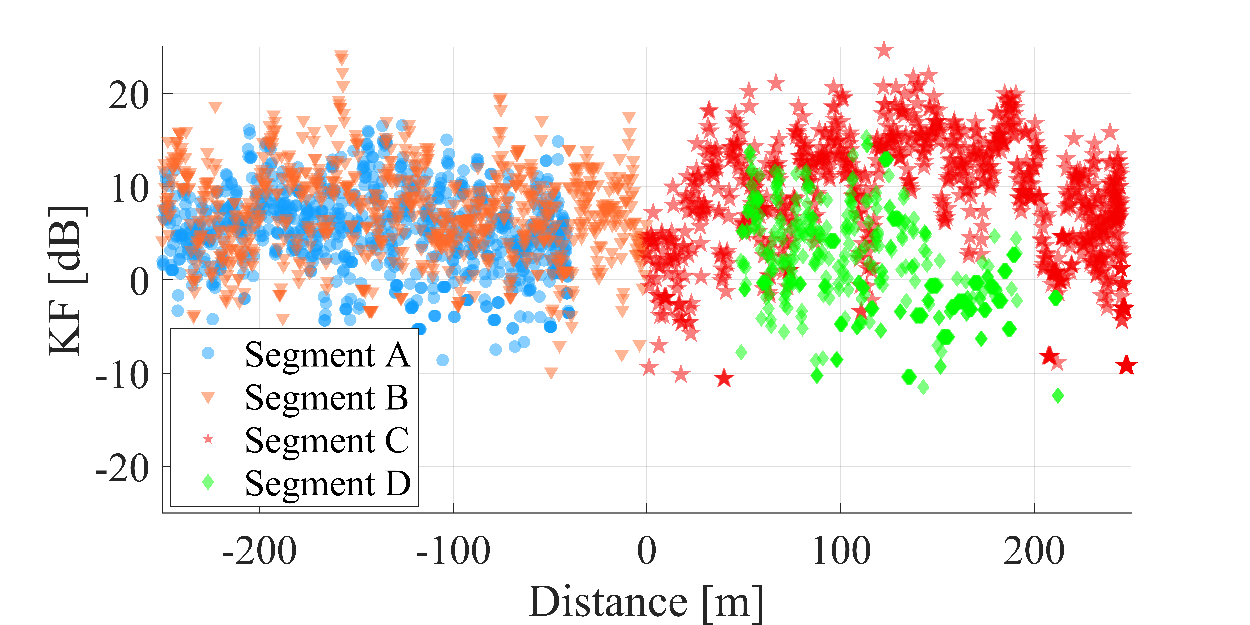}
}
\quad
\subfloat[]{
\includegraphics[scale=0.32]{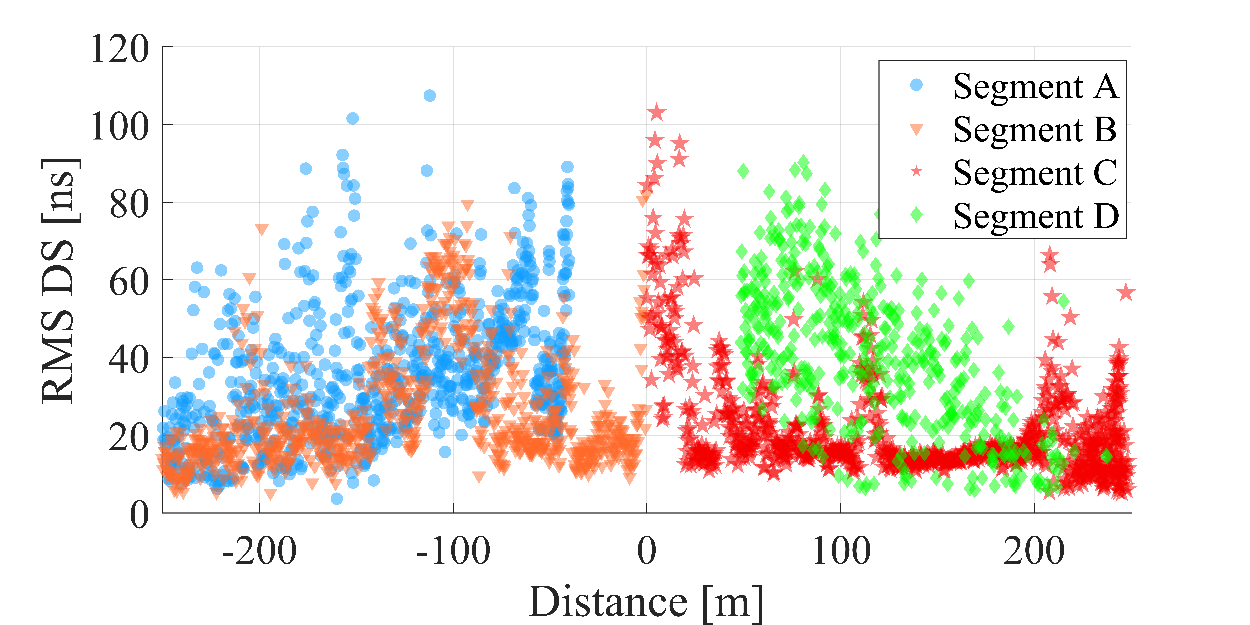}
}
\hspace{-0.2cm}
\subfloat[]{
\includegraphics[scale=0.32]{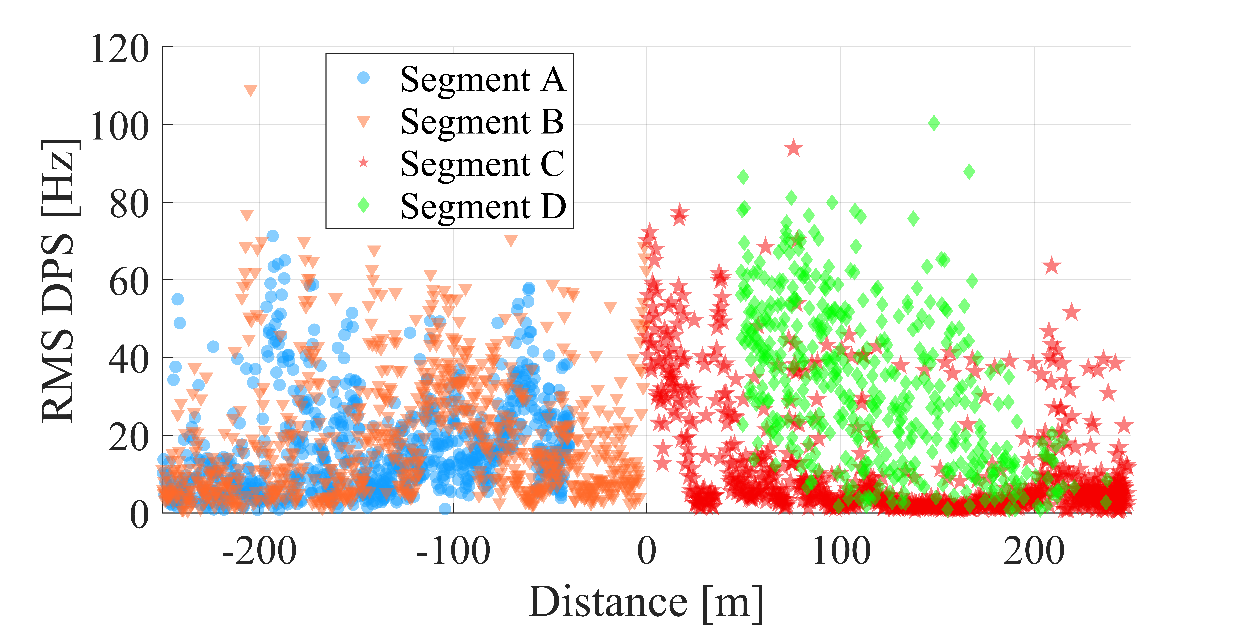}
}
\caption{Measurement results. (a) Number of MPCs. (b) KF. (c) RMS DS. (d) RMS DPS. \label{fig:16}}
\vspace{-0.45cm}
\end{figure*}
The CSF provides an intuitive visual representation of the channel in the DD domain. Fig. 14 illustrates two CSF examples measured at a T-R distance of 55 m, corresponding to the LOS and NLOS propagation, respectively. Due to the presence of fractional delay and Doppler effects, the CSF exhibits power leakage along both the delay and Doppler dimensions. Consequently, coarse channel parameters, which are constrained to integer multiples of the resolution, can be obtained from the measured CSF. By applying the proposed joint fractional delay and Doppler shift estimation algorithm to the measured CSF, refined channel parameters can be obtained. As an illustrative example, Figs. 14(c) and (d) present the refined results when the delay and Doppler steps are set to 0.1 and 0.01, which correspond to a 1 ns delay resolution and a 1.91 Hz Doppler resolution, respectively. In the LOS case, the instantaneous velocity is approximately 42 km/h, which corresponds to a theoretical maximum Doppler shift of 130.28 Hz. The CSF initially yields a coarse Doppler shift estimate of 190.73 Hz for the strongest path, which is an integer multiple of the original Doppler resolution. According to the refined estimation results, the Doppler shift of this path is 128 Hz, which aligns more closely with the theoretical value. Considering the angle between the arriving wave and the direction of motion, the actual Doppler shift is smaller than the theoretical value. Furthermore, the MPC distributions differ markedly between LOS and NLOS cases. {Considering MPCs with normalized powers above $-$20 dB, the LOS case contains only 8 MPCs, which are mainly concentrated around a Doppler shift of 128 Hz, whereas the NLOS case contains 24 MPCs distributed over a wider Doppler shift range from approximately $-$200 Hz to 200 Hz.}

The time-variant PDPs and DPSDs, both before and after applying the proposed joint fractional delay and Doppler shift estimation algorithm, are illustrated in Fig. 15. Fig. 15(a) depicts the PDP without refined estimation, wherein the trajectory of the strongest MPC clearly mirrors the variation in the actual T-R distance. Initially, the propagation delay increases as the TX moves away from the RX. After the first U-turn, the TX begins approaching the RX, causing a continuous decrease in the propagation delay while the received power increases significantly. Subsequently, the propagation delay gradually rises again as the TX recedes from the RX. Following the second U-turn, the TX is redirected back toward the RX, leading to a continuous decrease in delay until the measurement concludes. Fig. 15(b) presents the time-variant DPSD without refined estimation, where the evolution of the Doppler shift is governed by both velocity variations and the angle between the incident wave and the direction of motion. Initially, as the TX accelerates away from the RX, the Doppler shift starts at 0 Hz and continuously extends into the negative regime. Following the first U-turn, the TX approaches and eventually passes the RX, causing the Doppler shift to jump to a positive value, decrease to 0 Hz at the passing point, and then rapidly transition back to negative values as the TX recedes. After the second U-turn, the Doppler shift gradually decays from positive values toward 0 Hz. Unfortunately, constrained by the fractional Doppler effect, this behavior is only coarsely captured in the DPSD.
The implementation of the joint fractional delay and Doppler shift estimation algorithm across all measured CSFs enables precise extraction of time-varying parameters of MPCs. This results in enhanced PDP and DPSD representations, as shown in Figs. 15(c) and (d). The estimation procedure is configured with a 1 ns delay resolution and a 1.91 Hz Doppler resolution. {The maximum number of extracted MPCs is set to 60, while the dynamic power threshold is determined as 5 dB above the maximum noise floor level. The value of 60 is selected based on the maximum number of MPCs observed in the collected measurement data, with an additional margin to avoid prematurely limiting MPC extraction.} 
After refined estimation, the time-varying Doppler shifts are resolved with higher precision, revealing smoother transitions within a reduced uncertainty range compared to the unrefined measurements.

To further characterize the measured wireless channel, in addition to the aforementioned measurement results, we also present the number of MPCs, KF, RMS DS, and RMS DPS. 
Fig. 16 presents the variations of the number of MPCs, KF, RMS DS, and RMS DPS with respect to the T-R distance across the four segments. As observed, compared to segment C, segments A and D exhibit a larger number of MPCs, a lower KF, and noticeably higher RMS DS and RMS DPS. Furthermore, segment B at large T-R distances exhibits parameter variations similar to those in segment A, whereas its characteristic parameters at short T-R distances behave similarly to those in segment C. Fundamentally, segments A and D experience OLOS and NLOS propagation due to blockage from vegetation and the soundproof wall, respectively. Similarly, the OLOS condition in segment B at large T-R distances is caused by vegetation blockage introduced by road curvature, whereas both segment C and segment B at short T-R distances maintain well-preserved LOS environments. When the T-R distance approaches 0 m, the bridge structure supporting the RX induces substantial obstructive scattering, leading to a significant increase in the number of MPCs, a sharp reduction in the KF, and a rise in both the RMS DS and RMS DPS.

\subsection{{{V2V Channel Measurements}}}
\subsubsection{{{Measurement Campaigns}}}
{During the V2V measurements, the TX and RX devices are installed in two private cars, respectively, with both antennas mounted on the vehicle roofs, as shown in Fig. 17(a). The hardware configurations of both the TX and RX are the same as those used in the V2I measurements, and the RX devices installed inside the RX vehicle are shown in Fig. 17(b). The other measurement settings are also kept the same as in the V2I measurements.}

\begin{figure}[!t]
{
\center
\includegraphics[scale=0.25]{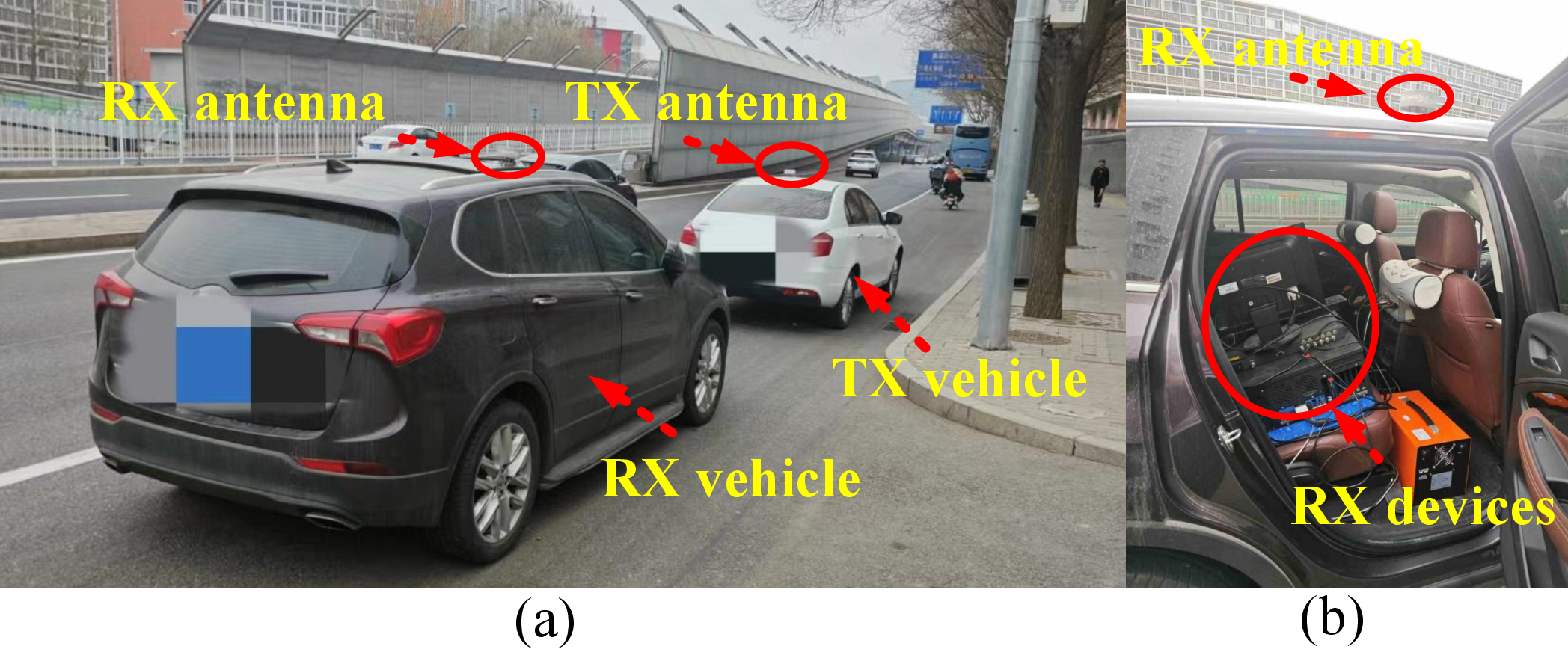}\\
\setlength{\abovecaptionskip}{-2pt}
\caption{V2V measurement equipment. (a) TX and RX vehicles. (b) RX devices and antenna. \label{fig:19}}
\vspace{-0.3cm}
}
\end{figure}

\begin{figure}[!t]
{
\center
\includegraphics[scale=0.28]{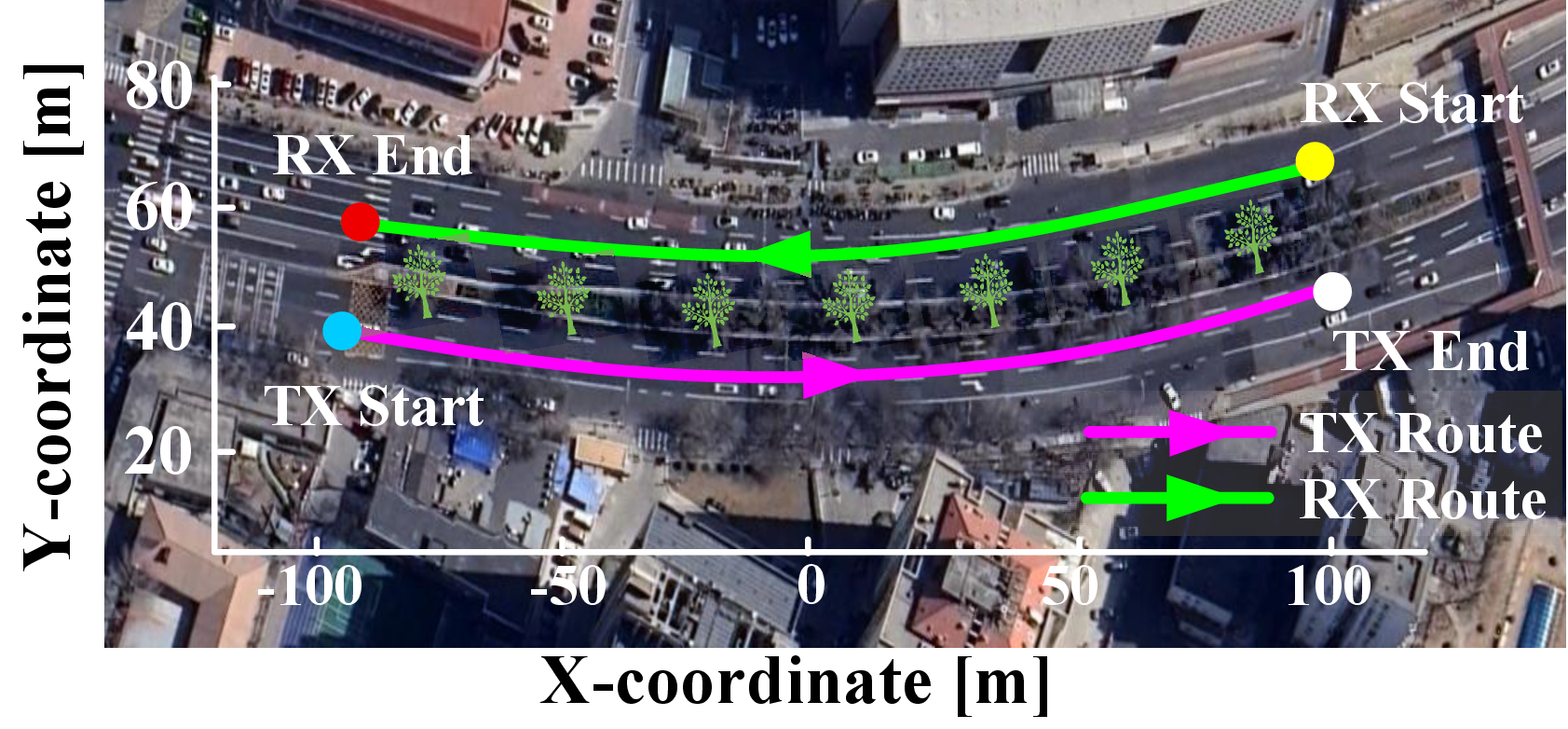}\\
\setlength{\abovecaptionskip}{-2pt}
\caption{{V2V measurement route.} \label{fig:20}}
\vspace{-0.3cm}}
\end{figure}

{The TX and RX vehicles travel in opposite directions along their respective routes, resulting in a relative driving speed of approximately 110 km/h. The V2V measurement is conducted in the same environment as the V2I measurement, and the corresponding TX and RX routes are shown in Fig. 18. During the measurement, vegetation in the central green belt partially obstructs the propagation between the TX and RX. The T-R distance varies from $-$60 m to 0 m and then to 60 m, where a negative sign indicates the period before the two vehicles encounter each other, 0 m corresponds to the encounter position, and a positive distance denotes the period after the encounter.}

\subsubsection{{{Measurement Results}}}
{The time-variant PDP and DPSD after applying the joint fractional delay and Doppler shift estimation algorithm during the V2V measurement are shown in Fig. 19. As the TX and RX vehicles approach each other and then move apart after the encounter, the delay of the dominant path first decreases and then increases, while its power generally increases and subsequently decreases. The DPSD exhibits stronger Doppler variation than that observed in the V2I measurement. Specifically, the Doppler shift of the dominant path evolves from approximately 320 Hz before the encounter to around 0 Hz during the encounter and then to approximately $-$300 Hz after the encounter, clearly reflecting the transition of the relative motion from approaching to receding. The observed maximum Doppler shift of approximately 320 Hz is close to the theoretical value of approximately 341 Hz corresponding to the relative driving speed of 110 km/h and is substantially larger than that observed in the V2I measurement.}

\begin{figure}[!t]
{
\centering
\subfloat[]{
\includegraphics[scale=0.33]{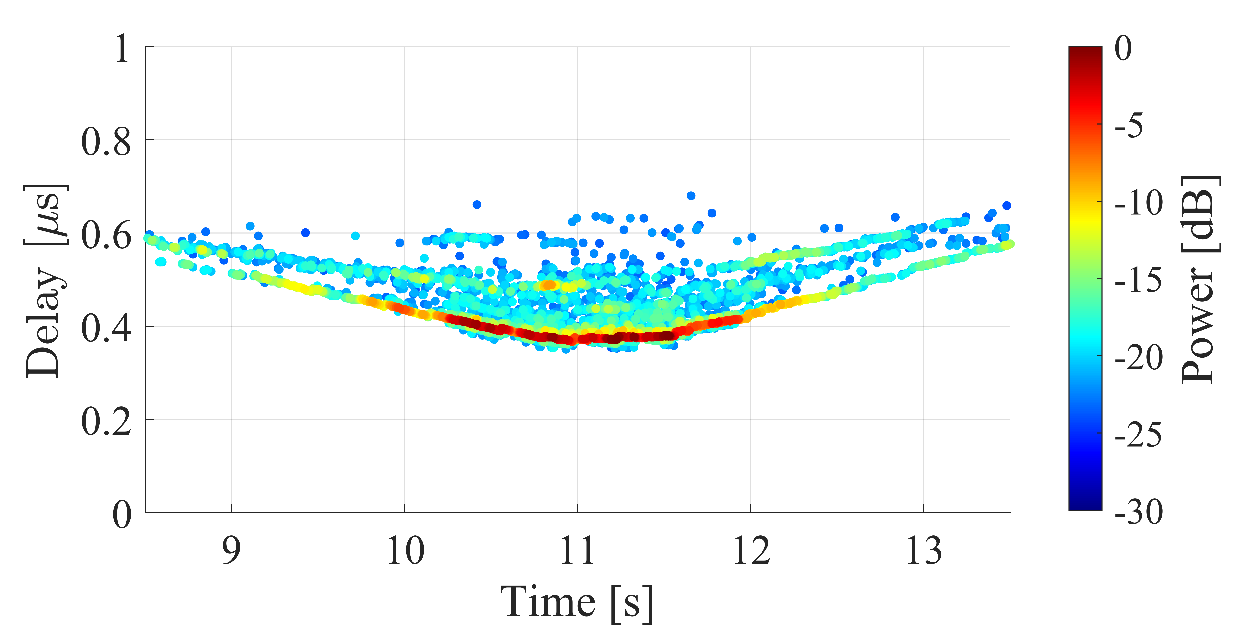}
}
\quad
\subfloat[]{
\includegraphics[scale=0.33]{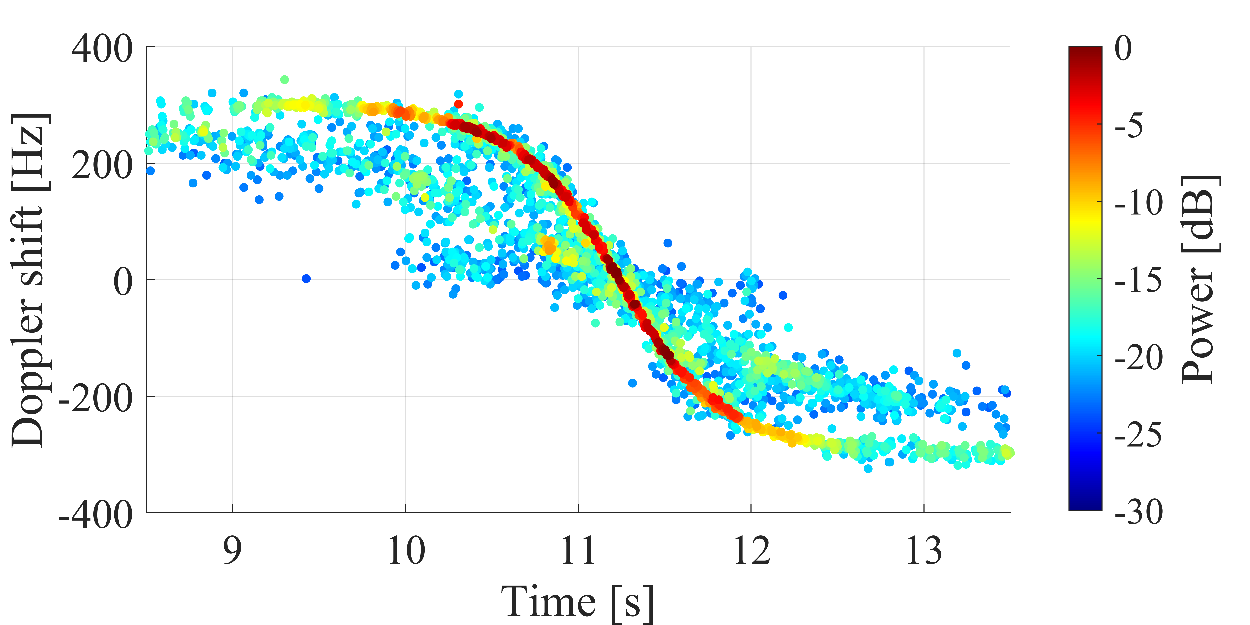}
}
\caption{Measured time-variant PDP and DPSD after joint fractional delay and Doppler shift estimation during the V2V measurement. (a) PDP. (b) DPSD. \label{fig:21}}
}
\vspace{-0.3cm}
\end{figure}

\begin{figure}[!t]
{
\centering
\subfloat[]{
\includegraphics[scale=0.34]{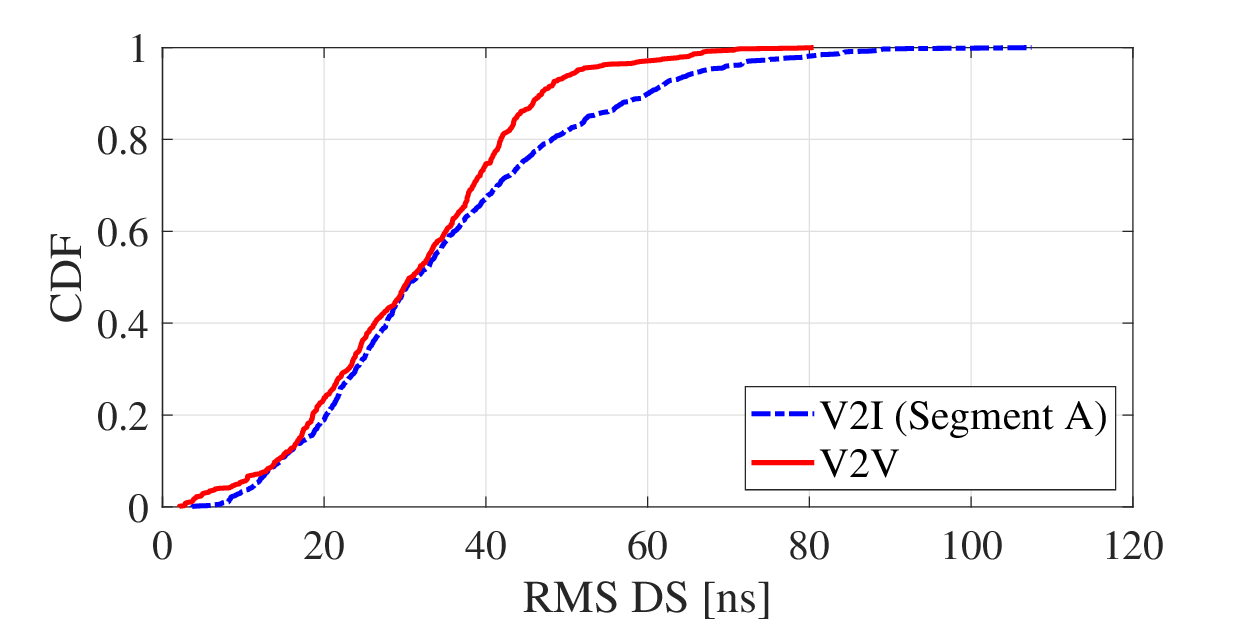}
}
\quad
\subfloat[]{
\includegraphics[scale=0.34]{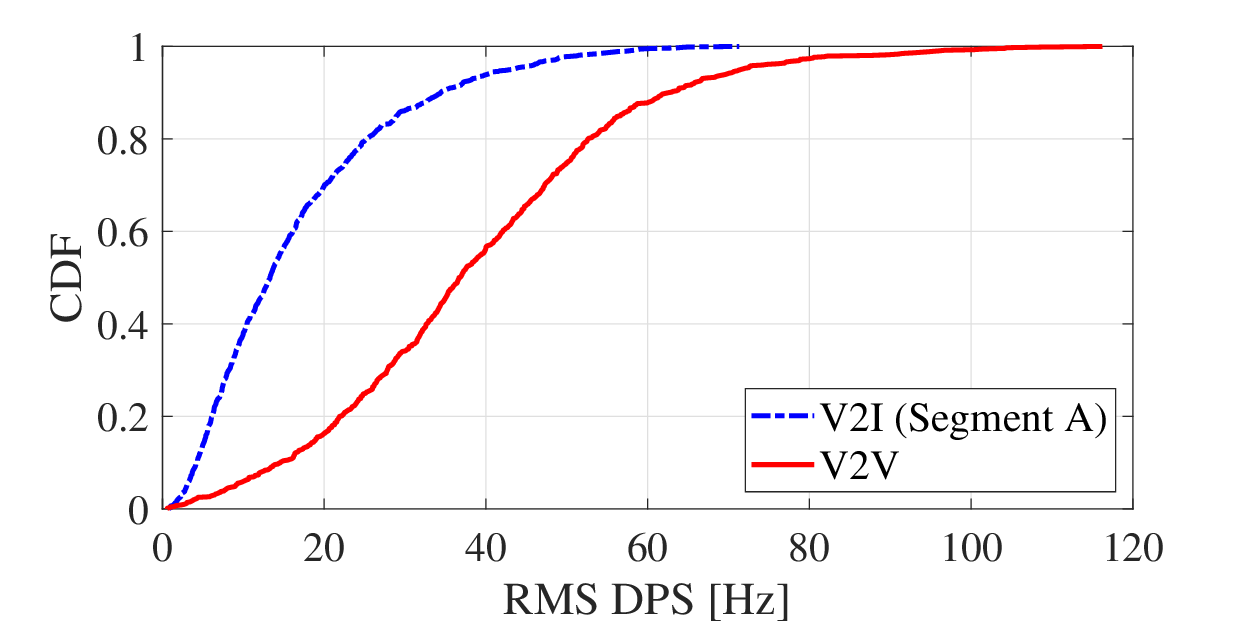}
}
\caption{Measurement results. (a) RMS DS. (b) RMS DPS. \label{fig:22}}
\vspace{-0.45cm}
}
\end{figure}

{The cumulative distribution functions (CDFs) of the RMS DS and RMS DPS obtained from the V2V measurement are further shown in Fig. 20. For comparison, the corresponding results for Segment A of the V2I measurement are also included, since its propagation environment is similar to that of the V2V measurement. The two scenarios exhibit similar RMS DS values, indicating comparable delay dispersion characteristics. In contrast, the V2V measurement exhibits a noticeably larger RMS DPS than V2I Segment A. This difference is mainly attributed to the higher relative mobility between the TX and RX and to MPCs arriving from different propagation directions, which together result in a broader Doppler distribution.}

\section{Conclusion}
In this paper, we have presented an OTFS waveform-based DD domain channel measurement method for high-mobility scenarios. By considering synchronization and PAPR performance, we have innovatively incorporated PN sequences into the waveform design of the sounding signal and have provided a comprehensive analysis of its sounding capability. The complete methodology for DD domain channel measurement, including synchronization and CSF estimation, has been thoroughly described. Specifically, a joint fractional delay and Doppler shift estimation algorithm for improving channel measurement accuracy has been proposed. {The performance of the proposed method has been rigorously evaluated in terms of PAPR, synchronization, CSF estimation, and fractional delay and Doppler shift estimation.} Furthermore, we have developed a practical DD domain channel measurement system. Following system calibration and comprehensive verification, the proposed method has been confirmed to be fully capable of supporting measurement campaigns in high-mobility scenarios. {Based on this system, DD domain channel measurements have been conducted in V2X urban scenarios, covering V2I and V2V scenarios.} Finally, the measurement results, including the CSF, PDP, DPSD, number of MPCs, KF, RMS DS and RMS DPS, have been derived, demonstrating the effectiveness of the proposed method and providing valuable insights for the design of next generation communication systems in high-mobility scenarios.


\newpage

\end{document}